\documentstyle[prd,tighten,aps]{revtex}
\newcommand{\alfabet}{
\renewcommand{\theequation}{\Alph{section}.\arabic{equation}}}

\begin{document}
\draft

\twocolumn[\hsize\textwidth\columnwidth\hsize\csname
@twocolumnfalse\endcsname
\title{\bf Quantum black-hole kinks}

\author{Pedro F. Gonz\'alez-D\'{\i}az}
\address{Centro de F\'{\i}sica "Miguel Catal\'an",
Instituto de Matem\'aticas y F\'{\i}sica Fundamental,\\
Consejo Superior de Investigaciones Cient\'{\i}ficas,
Serrano 121, 28006 Madrid (SPAIN)}
\date{March 10, 1996}

\maketitle

\begin{abstract}

By allowing the light cones to tip over on hypersurfaces
according to the conservation laws of an one-kink in static,
Schwarzschild and five-dimensional black hole metrics, we
show that in the quantum regime
there also exist instantons whose finite imaginary action
gives the probability of occurrence of the kink metric
corresponding to single 
chargeless, nonrotating black holes taking place in pairs,
joined on an interior surface, beyond the
horizon, with each hole residing in a different universe. 
Evaluation of the thermal properties of each of
the black holes in a neutral pair leads one to check that, to
an asymptotic observer in either universe,  
each black hole is exactly the quantum-mechanically
defined {\it anti-black hole} to the other
hole in the pair. The independent quantum states of black holes
in neutral pairs
have been formulated by using the path integral method, and
shown to be that of a harmonic oscillator. Our results
suggest that the boundary condition of a single universe 
in the metauniverse is
that this universe can never be self-contained and must
always have at least one boundary which connects it to the rest
of a self-contained metauniverse.

\end{abstract}
\pacs{PACS number(s): 04.70.Bw, 04.70.Dy, 04.60.Gw }

\vskip2pc]

\renewcommand{\theequation}{\arabic{section}.\arabic{equation}}

\section{Introduction}

The success of the thermodynamical analogy in black hole physics
allows us to hope that this analogy may be even deeper, and that
it is possible to develop a statistical-mechanical foundation
of black hole thermodynamics [1]. However, it is not quite clear
that black hole entropy may count the number of internal
degrees of freedom. These would describe different internal
states which may exist for the same values of the black hole
external parameters. It might be [2] that in the state of
thermal equilibrium the parameters for the internal degrees
of freedom will depend on the temperature of the system in
the universal way, thus cancelling all contributions which
depend on the particular properties and number of internal
fields.

Clearly, the most compelling argument in favour of the idea
that black hole entropy counts the number of internal states
has recently come from proposals of black hole pair creation [3-7].
By computing the exact action of the black hole pair instanton
it has been shown [4] that the action for the case of nonextreme
Reissner-Nordstrom black holes is smaller than that of the
corresponding pair creation rate by exactly a factor of
the black hole entropy $S_{BH}$; that is precisely what one
would expect if black holes had $e^{S_{BH}}$ internal states.
This is not the case nevertheless for extreme
Reissner-Nordstrom black holes where [5] the instanton action
exactly equals the rate of pair production, though this
apparent contradiction has been explained by the argument
that these black holes have zero entropy [6].

Apart from the fact that all these considerations are
based only on the leading order semiclassical approximation
and some higher order terms might be expected [7] to 
contribute importantly as well, the above analyses show two
limitations. First of all, they are restricted to deal
with Reissner-Nordstrom black holes. It would 
be of most interest if we could make a similar
analysis in Schwarzschild black holes, where there
is just an event horizon with unambiguous classical
localization and one external parameter, {\it i.e.}
the mass $M$. On the other hand, there seems to be
no clear idea about where exactly the internal states
may reside. They might be either inside the black hole
or on the horizon and, in the event the first possibility
applies, one may still wonder which region of the black
hole interior does form the geometrical domain for states,
so as exactly on what internal surface are identified the two black
holes in a pair while preserving the appropriate contribution
of the spacetime to the quantum construct.

The formation of neutral black-hole pairs with the two
holes residing in the same universe is believed to be a
highly suppressed process [8]. However, 
it has been suggested [9] that single black holes could
evaporate completely, taking with them all the particles that
fell in to form the black hole and the antiparticles to
emit radiation, by going off through a wormhole whose other
end opens up in other universe, forming another 
evaporating black hole which can be regarded to have been
formed from the collapse of exactly the massive antifermions
to the fermions that went in to form the first hole. One
could thus regard each single Schwarzschild black hole
in our universe as just a member of a neutral pair, with
the other hole of the pair residing in other universe.
In order to investigate the nature of such interuniverse
pairs, it appears
important to know what is the quantum state of
the wormhole that connects the two holes. 
If the wormhole is in a pure quantum state [10],
the universe should be self-contained with no real boundary
through which information may escape from it to
any other possible universes. If, on the contrary,
wormholes are in mixed quantum states [11], then one would not expect
the universe to be self-contained and its quantum state
could not be factored out from the rest of the metauniverse.

This paper deals with these issues by considering
the geometrical situation that results from regarding neutral
black holes as gravitational topological defects which
can move in spacetime but cannot be removed without cutting [12],
when such defects are viewed as quantum geometrical constructs.
As a consequence from imposing the kink metric in standard form
to hold along the entire radial-coordinate interval from
$\infty$ to some interior nonzero surface, and
invariance of the number
of the resulting D-dimensional black hole kinks, we can
obtain maximally-extended black hole kink metrics which
can only be described by means of two coordinate patches
for each one-kink. This extended kink metric can only
be consistently defined in some quantum realm and
represents two black holes, rather than one, with each
hole being described in just one patch and
the two patches identified just on the minimal single interior
surface [13]. If we interpret each coordinate patch as the
set of coordinates with which one just describes a single,
distinct universe, to an observer in one of the
asymptotically flat regions, the essential quantum
nature of the black-hole kink interior would manifest
in such a way that the observer can regard the whole
of the kink geometry like though if it were a neutral
black-hole pair, with each hole in the pair residing
in a distinct universe.
The two black holes of this pair are joined to each other at a
given interior surface through a wormhole thinner than
the corresponding Einstein-Rosen bridge [14].

We review D-dimensional black hole kinks and their 
connection to (D-1)-wormholes in Sec. II, and
discuss the black-hole kink instantons that describe
creation and annihilation of black-hole kinks in
vacuum in Sec. III.
Sec.IV deals with the quantum theory of black-hole
kinks by using the
Euclidean path integral approach. 
The probability that neutral black-hole pairs can occur
as a consequence from the conservation of the
topological charge of the kink is discussed in Sec. V,
where the thermal emission of such pairs is also
discussed, together with their quantization using both
the Euclidean path integral approach and a generalized
quantum theory in terms of a decoherence functional.
It is obtained
that single black holes behave like quantum harmonic
oscillators whose frequency corresponds to the minimum
possible internal energy compatible with the quantum
De Broglie relation between energy and wavelength.
A brief discussion
on the possible cosmological implications of our results
is included in Sec. VI, where a generalized boundary
condition for a universe which is not self-contained
is proposed. We summarize the main results in Sec. VII.
Finally, an Appendix is added which
contains a calculation of the action for the Schwarzschild
pairs. This action turns out to be smaller than that
of the corresponding rate of pair creation by the
Bekenstein-Hawking entropy factor.
Throughout the paper natural units so that $c=G=k_{B}=\hbar$
are used.

\section{\bf The Black-Hole Kinks}
\setcounter{equation}{0}

The idea that we shall explore in this section is based on looking at
the spacetime metric of a (D-1)-wormhole as the metric that results
on the constant-time hypersurfaces corresponding to purely future
directed or purely past directed light-cone orientations of a
D-dimensional black-hole spacetime where we allow all possible
light cone orientations compatible with the existence of a
gravitational kink [15]. We shall restrict to the physically most
interesting examples with D=4, which associated a Schwarzschild
black hole to a three-dimensional wormhole, and with D=5, which
corresponded to a five-dimensional Tangherlini black hole
associated with a four-dimensional Tolman-Hawking wormhole [15].

We first briefly review the general topological concept
of a kink and its associated topological charge.
Let $({\bf M},g_{ab})$ be a given D-dimensional spacetime, with
$g_{ab}$ a Lorentz metric on it. One can always regard $g_{ab}$
as a map from any connected D-1 submanifold $\Sigma\subset {\bf M}$
into a set of timelike directions in ${\bf M}$ [16]. Metric homotopy
can then be classified by the degree of this map. This is
seen by introducing a unit line field $\{n,-n\}$, normal to
$\Sigma$, and a global framing $u_{i}$: $i$=1,2,...,D-1, of
$\Sigma$. A timelike vector ${\bf v}$ can then be written in terms
of the resulting tetrad framing $(n,u_{i})$ as
$v=v^{0}n+v^{i}u_{i}$, such that $\sum_{i}^{D-1}(v^{i})^{2}=1$.
Restricting to time orientable manifolds ${\bf M}$, ${\bf v}$ then
determines a map
\[K: \Sigma\rightarrow S^{D-1}\]
by assigning to each point of $\Sigma$ the direction that
${\bf v}$ points to at that point. This mapping allows a general
definition of kink and kink number. Respect to hypersurface
$\Sigma$, the kink number (or topological charge) of the
Lorentz metric $g_{ab}$ is defined by [16]
\[kink(\Sigma;g_{ab})=deg(K),\]
so this topological charge measures how many times the
light cones rotate all the way around as one moves
along $\Sigma$ [17].

In the case of an asymptotically flat spacetime the pair
$(\Sigma,g)$ will describe an asymptotically flat kink if
$kink(\Sigma;g)\neq 0$. All of the topological charge of
the kink in the metric $g$ is in this case confined to
some finite compact region [17]. Outside that region all
hypersurfaces $\Sigma$ are everywhere spacelike. For the
case of a spherically symmetric kink, to asymptotic
observers, the compact region containing all of the
topological charge coincides with the interior of either
a black hole when the light cones rotate away from the
observers (positive topological charge), or a white
hole when the asymptotic observers "see" light cones
rotating in the opposite direction, toward them (negative
topological charge).

Topology changes, such as handles or wormholes, can occur
in the compact region supporting the kink, but not outside
it. All topologies are actually allowed to happen in such
a region. Therefore, in the case of spherically symmetric
kinks, the supporting region should be viewed as an
essentially quantum-spacetime construct. This is the
view we shall assume throughout this paper.

\subsection{The Schwarzschild Kink}

We can take for the static, spherically symmetric metric of a
three-dimensional wormhole
\begin{equation}
ds^{2}=(1-\frac{2M}{r})^{-1}dr^{2}+r^{2}d\Omega_{2}^{2},
\end{equation}
where $d\Omega_{2}^{2}$ is the metric on the unit two-sphere.
Metric (2.1) describes a spacetime which (i) is free from any
curvature singularity at $r=0$, and (ii) possesses an apparent
(horizon) singularity at $r=2M$ that is removable by a suitable
coordinate transformation. The re-definition
\begin{equation}
r=\frac{M}{2}\left(\frac{u}{\mu}+\frac{\mu}{u}\right)^{2},
\end{equation}
where $\mu$ is an arbitrary scale, transforms metric (2.1) into
\begin{equation}
ds^{2}=\frac{M^{2}}{4}\left(\frac{u}{\mu}+\frac{\mu}{u}\right)^{4}\left(\frac{4du^{2}}{u^{2}}+d\Omega_{2}^{2}\right).
\end{equation}
Along the complete $u$-interval, ($\infty$,$\mu$), metric (2.3)
varies from an asymptotic region at $u=\infty$ to a minimum
throat at $u=\mu$ (i.e. at $r=2M$).

On the other hand, metric (2.1) is in fact a constant time
section, $T=t_{0}$, of a Schwarzschild black hole,
\begin{equation}
ds^{2}=-(1-\frac{2M}{r})dT^{2}+(1-\frac{2M}{r})^{-1}dr^{2}+r^{2}d\Omega_{2}^{2}.
\end{equation}
Metric (2.1) can likewise be regarded as being described by
constant Euclidean time $\tau=-iT$ sections of the
Gibbons-Hawking instanton [18] associated to (2.4). Each of
such three-wormholes would correspond to a given Einstein-Rosen
bridge on this instaton, so that one of the two halves of
the wormhole should then be described in the unphysical [19]
exterior region created in the Kruskal extension of metric
(2.4). In order to avoid the need of using such an unphysical
region to describe a complete wormhole,
we shall consider that metric (2.1) corresponds to a given fixed
value of time $T$ in the kink extension
of (2.4). 

We take for the metric that describes a spherically
symmetric one-kink in four dimensions [12],
\begin{equation}
ds^{2}=-\cos 2\alpha(dt^{2}-dr^{2})\pm 2\sin 2\alpha
dtdr+r^{2}d\Omega_{2}^{2}, 
\end{equation}
where $\alpha$ is the angle of tilt of the light cones, and
the choice of sign in the second term depends on whether a
positive (upper sign) or negative (lower sign) topological
charge is being considered. An one-kink is ensured to exist
if $\alpha$ is allowed to monotonously increase from 0 to
$\pi$, starting with $\alpha(0)=0$. Then metric (2.5)
converts into (2.4) if we use the substitution
\begin{equation}
\sin\alpha=\sqrt{\frac{M}{r}}
\end{equation}
and introduce a change of time variable $t+g(r)=T$, with
\begin{equation}
\frac{dg(r)}{dr}=\tan 2\alpha .
\end{equation}
Now, since $\sin\alpha$ cannot exceed unity, it follows that
$\infty\geq r\geq M$, so that $\alpha$ varies only from 0
to $\frac{\pi}{2}$. In order to have a complete one-kink
gravitational defect, we need therefore a second coordinate
patch  to describe the other half of the $\alpha$ inteval,
$\frac{\pi}{2}\leq\alpha\leq\pi$.

The kink metric (2.5), which is defined by coordinates $t$, $r$,
$\theta$,$\phi$ and satisfy (2.6) and (2.7), restricts the
Schwarzschild solution to cover only the region
$\infty\geq r\geq M$. If one wants to extend such a metric
to describe the region beyond $r=M$ as well, two procedures
can in principle be followed: (i) if the compact support of
the kink is assumed to be classical, then one lets $\alpha$
continue to increase as $r$ decreases from $r=M$ until 
$\alpha=\pi$ at $r=0$ to produce a manifold which has a
homotopically nontrivial light cone field and one kink.
This procedure makes metric (2.5) and definitions (2.6) and (2.7)
to hold asymptotically only, and since, classically, one
should assume a continuous distribution of matter in the
kink support, the momentum-energy tensor can be chosen to
satisfy reasonable physical conditions such as the weak
energy condition [12]. (ii) The second procedure 
can apply when
one assumes the black hole interior (i.e. the supporting
compact region of the kink) to be governed by quantum mechanics.
The simplest quantum condition to be satisfied by the
interior region supporting a black- or white-kink arises from
imposing metric (2.5) to hold along the radial
coordinate interval of the kink, i.e.: $\infty\geq r\geq M$,
rather than asymptotically only. Actually, in this case,
the kink geometry
should hold in the two coordinate patches which we need to
describe the complete one-kink gravitational defect. The
need for a second coordinate patch can most clearly be seen
by introducing the new time coordinate
\begin{equation}
\bar{t}=t+h(r),
\end{equation}
which transforms metric (2.5) into the standard metric [12]
\begin{equation}
ds^{2}=-\cos 2\alpha d\bar{t}^{2}\mp 2kd\bar{t}dr+r^{2}d\Omega^{2}_{2},
\end{equation}
provided
\begin{equation}
\frac{dh(r)}{dr}=\frac{dg(r)}{dr}-\frac{k}{\cos 2\alpha},
\end{equation}
with $k=\pm 1$ and the choice of sign in the second term
of (2.9) again depending on whether a positive (upper sign)
or negative (lower sign) topological charge is considered.
The choice of sign in (2.10) is adopted for
the following reason. The zeros of the denominator of
$dh/dr=(\sin 2\alpha\mp 1)/\cos 2\alpha$ correspond to
the two horizons where $r=2M$, one per patch. For the first
patch, the horizon occurs at $\alpha=\frac{\pi}{4}$ and
therefore the upper sign is selected so that both $dh/dr$
and $h$ remain well defined and hence the kink is not lost
in the transformation from (2.5) to (2.9). For the second
patch the horizon occurs at $\alpha=\frac{3\pi}{4}$ and
therefore the lower sign in (2.10)
is selected. $k=+1$ will then
correspond to the first coordinate patch and $k=-1$ to
the second one.

Metric (2.9) can be transformed directly into the
Schwarzschild metric (2.4) if we use (2.6) and the new
coordinate transformation
\begin{equation}
\bar{t}=T-f(r),
\end{equation}
where
\begin{equation}
\frac{df(r)}{dr}=\frac{k}{\cos 2\alpha}.
\end{equation}

We impose then the standard kink metric (2.9) to hold along
$\infty\geq r\geq M$ on the two patches $k=\pm 1$. It will
be seen in Sec. VB that the holding of
this condition ultimately amounts to a black-hole
quantum interior whose energy spectrum is given by
$\frac{k}{M}(n+\frac{1}{2})$, $n=0,1,2,...$. The imposing
of a kinky metric like (2.9) also in the interior
of the hole (supporting region of the kink [17]),
together with the resulting harmonic-oscillator spectrum,
becomes a suitable quantization condition of such a
region for the following reasons. For the first
patch, the energy concentrated at $n=0$ on
the event horizon at $r=2M$ will correspond to a
pure vacuum energy, interpretable as a cosmological constant
which is, in any case, compatible 
with the holding at $r=2M$ of the kink metric (2.9), such as
it happens with the similar kinky metric at the cosmological
horizon in de Sitter space [20]. Substracting then the
zero-point energy corresponding to $n=0$,
along the radial coordinate
interval $2M>r>M$ there is
no spherical surface with nonzero energy and therefore this
interval does not contribute the stress tensor $T_{\mu\nu}$.
As one gets at $r=M$ on the first patch it would appear
a delta-function-like concentration of positive energy on
that surface corresponding to the quantum level $n=1$. This
would at first glance blatantly violate energy-momentum
conservation. However, the continuity of the angle of
tilt $\alpha$ at $\frac{\pi}{2}$ implies that the two
coordinate patches are identified at exactly the surfaces
$r=M$. Thus, since there would be an identical delta-function-like
concentration of negative energy-momentum at 
$n=1$ on $r=M$ in the
second patch, the total stress tensor $T_{\mu\nu}$ will also be
zero at the minimal surface $r=M$. Thus, although the
Birkhoff's theorem ensures [21] the usual Schwarzschild metric
as the unique spherically symmetric solution to the
four-dimensional vacuum Einstein equation, the violation
of this classical result the way we have shown above implies
an allowed {\it quantized} extension from it because this
extension entails no violation of energy-momentum conservation at any
interior spacelike hypersurface.

This result should be interpreted as follows. All what is
left at length scales equal or smaller than the minimum
size of the bridge (i.e. for $n\geq 1$, $r\leq M$) is some
sort of quantized "closed" baby universe with maximum size
$M$, whose zero total energy may be regarded as the sum of
the opposite-sign eigenergies of two otherwise identical
harmonic oscillators with the zero-point energy substracted.
The positive energy oscillator would play the role of the
matter field part of a constrined Hamiltonian, $H=0$, and
the negative energy oscillator would behave like though it
were the gravitational part of this Hamiltonian
constraint. On the
other hand, to an asymptotic observer in either patch, the above
quantized kink geometry would look like that of a black hole if the
topological charge is positive, and like that of a white hole if
the topological charge is negative. In the latter case,
to the asymptotic observer there would actually be
a topological change by which 
an asymptotically flat space converts into asymptotically
flat space plus a baby universe being branched off from it.
Now, since from a quantum-mechanical standpoint white and
black holes with the same mass are physically indistinguishable [22],
it follows that to asymptotic observers 
the asymptotically flat space of black holes is
physically indistinguishable from 
asymptotically flat space plus a baby
universe, with such a baby universe living outside the
realm of the two coordinate patches where the kink is
defined, in the inaccessible region between $r=M$ and
$r=0$.

Metric (2.9) contains still the geodesic incompleteness at
$r=2M$ of metric (2.4). This incompleteness can be removed by
the use of Kruskal technique. Thus, introducing the metric
\begin{equation}
ds^{2}=-F(U,V)dUdV+r^{2}d\Omega_{2}^{2},
\end{equation}
in which
\begin{equation}
F=\frac{4M\cos 2\alpha}{\beta}\exp\left(-2\beta
k\int^{r}_{\infty/M}\frac{dr}{\cos 2\alpha}\right),
\end{equation}
\begin{equation}
U=\mp e^{\beta\bar{t}}\exp\left(2\beta
k\int^{r}_{\infty/M}\frac{dr}{\cos 2\alpha}\right),
\end{equation}
\begin{equation}
V=\mp\frac{1}{2\beta M}e^{-\beta\bar{t}},
\end{equation}
where $\beta$ is an adjustable parameter which will be chosen
so that the unphysical singularity at $r=2M$ is removed,
and the lower integration limit $\infty/M$ refers to the
choices $r=\infty$ and $r=M$, depending on whether the
first or second patch is being considered. Using (2.6)
we obtain from (2.14)
\[F=4M\left(\frac{1-\frac{2M}{r}}{\beta}\right)\left(\frac{r}{M}\left(\frac{2M}{r}-1\right)\right)^{-4\beta
kr}.\]
This expression would actually have some constant term
coming from the lower integration limits $\infty/M$. We
have omitted at the moment such a term because it is
canceled by the similar constant term which appears
in the Kruskal coordinate $U$ when forming the Kruskal metric
from (2.13)-(2.16).

Unphysical singularities are then avoided if we choose
\begin{equation}
\beta=\frac{1}{4kM} .
\end{equation}
Whence
\begin{equation}
F=\frac{16kM^{3}}{r}e^{-\frac{r}{2M}},
\end{equation}
\begin{equation}
U=\mp e^{\frac{\bar{t}}{4kM}}e^{\frac{r}{2M}}(\frac{2M-r}{M}),
\; \; \; V=\mp\frac{k}{2}e^{-\frac{\bar{t}}{4kM}},
\end{equation}
where [15]
\[\bar{t}=t_{0}-k\int_{\infty/M}\frac{dr}{\cos 2\alpha}\]
\begin{equation}
=\bar{t}_{0}-k\left(r-2M\ln\left(\frac{M}{2M-r}\right)\right),
\end{equation}
with the constant $\bar{t}_{0}$ being obtained from $t_{0}$ after
absorbing the term arising from the lower integration limit
$\infty$ or $M$, depending on whether the first or second patch
is being considered. We finally obtain for the Kruskal metric
of the Schwarzschild kink
\begin{equation}
ds^{2}=-\frac{32kM^{3}}{r}e^{-\frac{r}{2M}}dUdV+r^{2}d\Omega_{2}^{2}.
\end{equation}
Except for the sign parameter $k$, this metric is the same as the
Schwarzschild-Kruskal metric.

Because of continuity of the angle of tilt $\alpha$ at $\frac{\pi}{2}$,
the two coordinate patches can be identified to each other only
on the surfaces at $r=M$ [13,15]. Such an identification should occur
both on the original and the new regions created by the Kruskal
extension, and represents a bridge that connects asymptotically
flat regions of the two coordinate patches. Any $T=$const. section
of this spacetime construct will then describe
halves of a three-dimensional
wormhole whose neck is now at $r=M$, rather than $r=2M$.
One can then describe the two halves of a complete 
wormhole just in the physical
original regions of either patch $k=+1$ or patch $k=-1$.

The causal structure of the considered geometry could at first
glance be thought of as being unstable due to mass-inflation
caused by the unavoidable presence of a Cauchy horizon [23]:
because quanta that enter the future event horizon at
arbitrary late time suffers an arbitrarily large blue shift
while propagating parallel to the Cauchy horizon, there will
be in general a mass-inflation singularity along a part of the
horizon in one patch caused by small fluctuations in the other.
However, using the spherical shell approach in the lightlike
limit [24] where a mass shell is allowed to move toward $r=0$
in the field of an interior mass distribution, it can be shown
that our kink model with quantized support prevents the
occurrence of any mass-inflation singularity. In fact, any
interior energy fluctuation in one patch is necessarily
sign-reversed to the energy of the imploding shell in the
other. Therefore, a mass increase must now occur in the 
expanding fluctuation shell, rather than in the imploding
shell, and the mass variation of these two shells is
nonsingular everywhere for $r\geq M$, even at the collision
radius where one would expect the mass singularity to
occur. We actually expect that, at that radius, imploding
and expanding gravitational masses are both finite with
half and twice their respective asymptotic values.

\subsection{The D=5 Tangherlini Kink}

The static, spherically symmetric metric of a four-dimensional
wormhole can be written as [25]
\begin{equation}
ds^{2}=(1-\frac{M^{2}}{r^{2}})^{-1}dr^{2}+r^{2}d\Omega_{2}^{2},
\end{equation}
where $d\Omega_{3}^{2}$ is the metric on the unit three-sphere.
As for the three-dimensional wormhole (2.1), we can regard
(2.22) as a singularity-free spacetime with removable apparent
singularity at $r=M$, describing a four-dimensional wormhole.
This is better seen by using the coordinate re-definition
\begin{equation}
r=\frac{M}{2}\left(\frac{v}{\nu}+\frac{\nu}{v}\right),
\end{equation}
in which again $\nu$ is an arbitrary scale measuring the size
of the wormhole throat. Using (2.23), (2.22) becomes
\begin{equation}
ds^{2}=\frac{M^{2}}{4}\left(\frac{v}{\nu}
+\frac{\nu}{v}\right)^{2}\left(\frac{dv^{2}}{v^{2}}+d\Omega_{3}^{2}\right).
\end{equation}
Metric (2.24) describes the connection between two asymptotically flat
regions through a four-dimensional wormhole with minimum throat at
$\nu$; i.e. at $r=M$. Now, we regard metric (2.22) as a constant
time section of the kink extension of the five-dimensional
Tangherlini black-hole metric [26]
\begin{equation}
ds^{2}=-\left(1-\frac{M^{2}}{r^{2}}\right)dT^{2}
+\left(1-\frac{M^{2}}{r^{2}}\right)^{-1}dr^{2}+r^{2}d\Omega_{3}^{2}.
\end{equation}
Also in this case metric (2.22) can be regarded as well as
being described by constant Euclidean time $\tau=-iT$ sections
of the five-dimensional black hole instanton, so that each of
such wormholes would correspond to a given
four-dimensional Einstein-Rosen
bridge on this instanton, with one of the two wormhole
halves necessarily being in the unphysical exterior region
created in the Kruskal extension of (2.25). Also in this
case, we shall avoid the need of using
this unphysical region for the description of a complete
wormhole by using the kink version of (2.25). The kink
extension in this case has been considered in detail
elsewhere [15]. Let us summarize the results in what
follows. By using arguments similar to those leading
to metric (2.9),
we can write the kink extension of (2.25) as [13,15]
\begin{equation}
ds^{2}=-\cos 2\alpha d\bar{t}^{2}\mp 2kd\bar{t}dr+r^{2}d\Omega_{3}^{2},
\end{equation}
with the choice of sign in the second term again referring
to whether a positive (upper sign) or negative (lower sign)
topological charge is being considered.

Metric (2.26) is nonstatic, requires two coordinate patches, $k=\pm 1$,
and can be obtained from (2.25) by using the functional relation
\[\sin\alpha=\frac{M}{\sqrt{2}r}\]
and the same time transformation as in (2.6) and (2.7). It also
contains a geodesic incompleteness at the horizon $r=M$ which
again is removable by the Kruskal technique. Imposing metric
(2.26) to hold along the interval $\infty\geq r\geq\frac{M}{\sqrt{2}}$
in the two patches and using arguments similar to those
in Sec.VB for a convenient definition of the propagator
through the corresponding complete wormhole, leads
again to a black hole spectrum that allows a consistent
quantum interpretation of the interior compact region supporting
the kink which is analogous to that for the four-dimensional
example considered in the precedent subsection.
In this case, the
Kruskal coordinate becomes finally [15]
\begin{equation}
U=\mp e^{\frac{\bar{t}}{kM}}e^{\frac{2r}{M}}\left(\frac{r-M}{r+M}\right),
\;\;\; V=\mp\frac{k}{2}e^{-\frac{\bar{t}}{kM}},
\end{equation}
with [15]
\begin{equation}
\bar{t}=\bar{t}_{0}-k\left(r+\frac{M}{2}\ln\left(\frac{r-M}{r+M}\right)\right),
\end{equation}
where the constant $\bar{t}_{0}$ is obtained from $t_{0}=T=$Const.
following the same procedure as in (2.20); i.e.: now by absorbing
into $t_{0}$ the respective constant term coming from the two
distinct integration lower limits $\infty$ and $\frac{M}{\sqrt{2}}$
corresponding to the two coordinate patches. The maximally-extended
metric for a Tangherlini five-dimensional kink becomes then
\begin{equation}
ds^{2}=-\frac{kM^{2}(r+M)^{2}}{r^{2}}e^{-\frac{2r}{M}}dUdV+r^{2}d\Omega_{3}^{2}.
\end{equation}
Again, this metric describes a spacetime which is nonsingular.
Continuity of the angle of tilt at $\alpha=\frac{\pi}{2}$ implies
the nonexistence of any surface with $r<\frac{M}{\sqrt{2}}$,
and the identification
of the surfaces at $r=\frac{M}{\sqrt{2}}$ of the two
coordinate patches, both on
the original regions described by metrics (2.26) and (2.29), and
on the new regions created by the the Kruskal extension. Thus,
one can continue the asymptotically flat spacetime described in
patch $k=+1$ into another asymptotically flat spacetime described
in patch $k=-1$ by means of a bridge at $r=\frac{M}{\sqrt{2}}$.
Every $T=t_{0}$
constant section of the resulting spacetime will thus represent
halves of
a Tolman-Hawking wormhole in four dimensions whose neck is now
at $r=\frac{M}{\sqrt{2}}$, rather than $r=M$. Thus, also
in this case, all wormhole halves can be described in the
physical regions of either patch $k=+1$ or patch $k=-1$.

\subsection{Travelling Through Wormholes}

If we replace the expression of $\bar{t}$ given by
(2.20) and (2.28) in the correponding Kruskal coordinates $U$,$V$,
we recover the usual expressions for these coordinates
outside the horizon [27] for both coordinate patches in
terms of $r$ and $\bar{t}_{0}$ (or its nonconstant
version $T$). In principle, one could therefore simply
continue in terms of these $U$,$V$ themselves inside the
horizon to obtain Kruskal coordinates which are real
everywhere. However, the need for two coordinate patches
(which continuously follow one another as the tilt angle
is monotonously varies from 0 to $\pi$) to describe an
one-kink makes it impossible to have coordinates $U$,$V$
which are real everywhere in the two patches while keeping
the original metric (2.4), or (2.25), Lorentzian on the
two patches. In fact, it is easy to check that the lower
integration limit one should use to get time $\bar{t}$
in one patch is necessarily different from the limit
that must be used at the same side of the horizon 
in the other patch. Since all that is physically required
is having a real nonsingular metric in the range
$\infty\geq r >0$, and the times $\bar{t}_{0}$ and
$T$ in (2.20) and (2.28) can by no means be kept real
in the same $r$-regions of the two patches simultaneously,
instead of continuing $U$,$V$, we shall continue time
$\bar{t}$ itself while keeping the Kruskal metric real
and expressions (2.19) and (2.27) for $U$,$V$ in terms of
$\bar{t}_{0}$ or $T$ valid everywhere. Different
continuations of time $\bar{t}$ that lead either to
some imaginary sectors for coordinates $U$,$V$, or to
instantonic transitions with $U$,$V$ real everywhere,
will respectively be discussed in what follows and in
Sec. III.

As it was pointed out before, the constant $\bar{t}_{0}$ is obtained
by adding to $t_{0}$ the distinct constants coming from the lower
integration limits in (2.20) and (2.28), and is the same in the
two coordinate patches [15]. 
If $\bar{t}_{0}$ would take on real values
in (2.20), i.e. if one considers metric (2.4) to be complex on
patch $k=+1$ and Lorentzian in patch $k=-1$, then though the
Kruskal metric (2.21) remains real and unchanged, coordinates
$U$, $V$ become both imaginary outside the black hole and both real
inside the black hole, in the two coordinate patches. By taking
into account that the lower integration limit in (2.20) gives rise
for $k=+1$ to a complex constant whose imaginary part is
$2\pi iM$, if we continue the chosen real value of $\bar{t}_{0}$
so that $\bar{t}_{0}\rightarrow\bar{t}_{0}-2\pi iM$, metric
would be Lorentzian on the first patch and complex in the
second one. On the two patches then coordinates $U$, $V$ would
be both real outside the horizon and both imaginary inside the
horizon, while the Kruskal metric (2.21) remained real and
unchanged. It follows that the spacetime description of
three-dimensional wormholes, each with neck at $r=M$
and two asymptotic physical regions,
requires two different four-dimensional black-hole spacetimes,
one in each patch, and therefore every complete three-dimensional
wormhole with neck at $r=M$, rather than $2M$,
cannot correspond to a single $T=$const. section of
the same Schwarzschild metric in the two patches.

As for the five-dimensional black hole, we obtain that real
values of $\bar{t}_{0}$ correspond to a black hole metric
(2.25) which is Lorentzian on patch $k=+1$ and complex on
patch $k=-1$. In this case, although also metric (2.29)
remains real and unchanged, coordinates $U$, $V$ are both
real for $r>M$ and imaginary beyond the horizon. If now
$\bar{t}_{0}$ is continued so that $\bar{t}_{0}\rightarrow
\bar{t}_{0}-\frac{\pi}{2}iM$, then metric (2.25) would be
complex for patch $k=+1$ and Lorentzian for patch $k=-1$,
with the Kruskal coordinates $U$, $V$ being real only for
$r\leq M$, and imaginary otherwise. Thus, the spacetime
description of a four-dimensional Tolman-Hawking wormhole
with neck at $r=\frac{M}{\sqrt{2}}$ and
two asymptotic physical regions also requires two different
five-dimensional black-hole spacetimes, one per patch, and
it is not possible to describe four-dimensional wormholes 
with neck at $r=\frac{M}{\sqrt{2}}$, rather than $M$, as
single $T=$const. sections of the same five-dimensional
Tangherlini black-hole metric in the two patches.

It may be instructive to look at null geodesics travelling
through the spacetime of these spacetimes from one
asymptotic region to the other (Fig. 1). Let us consider, for
example, the null geodesic which starts at spatial infinity
on the original region described by metric (2.4) on the
coordinate patch $k=+1$, crossing the horizon at $r=2M$ on
the $V$ axis to get in the original interior region. This
geodesic will then cross the identified surface at $r=M$
over into the original interior region of the second patch,
sharing on such a surface a common value of 
time $\bar{t}$ for the
two patches. Further propagation of the same geodesic first
crossing the $U$ axis of patch $k=-1$ to get in a new
(unphysical) exterior region to reach spatial infinity 
would at first sight seem to be
impossible as the crossing of the $U$ axis would take an infinite
time. Thus, the geodesic would seem to be trapped in the interior of
the black hole of patch $k=-1$. However, since the local
geometry is that of a Schwarzschild kink with standard
metric (2.9), or (2.26), everywhere even at $r=A$, it would
actually take a generally finite amount of proper time
with respect to such a metric for null geodesics to reach
the past event horizon ($\alpha=\frac{3\pi}{4}$) of the
second patch, as also it does entering the future event
horizon ($\alpha=\frac{\pi}{4}$) of the first patch.
The same effect will also be
obtained for null geodesics in the five-dimensional spacetime.
It will be shown in Sec. VA that the black hole in patch
$k=-1$ must necessarily have been formed from exactly the
antiparticles to the particles that went in to form the black
hole in patch $k=+1$. Thus, on the identified surface at
$r=A$ (with $A=M$ for D=4 and $A=\frac{M}{\sqrt{2}}$ for
D=5), the positive energy of the null ray would be instantly
annihilated, leading to a decrease of the absolute value of
the black hole mass. Balance between the two black hole masses
will be restored if, upon the arrival of the first ray at the
black hole in the second patch, this would emit {\it another}
ray with the same positive energy along an escaping path
parallel to the $U$ axis, into the original exterior region of the
black hole in the second patch. The whole process will be
equivalent to the simultaneous (in time $\bar{t}$) travel
of a null ray with positive energy from the original
spatial infinity in patch $k=+1$ and a null ray with
the same but negative energy from the original spatial
infinity in patch $k=-1$, both parallel to the $U$ axes,
which encounter and annihilate each other at the
surface $r=A$.

Inspection of Fig. 1 might sugget that null geodesics could
be closed timelike curves that somehow loop back through
the new regions created by the Kruskal extension. It is not
difficult to see however that this cannot be the case since,
though surfaces $r=A$ in the two patches are identified,
the identification of any surfaces at $r=\infty$ of the
two patches is disallowed, so null geodesics can never
complete a closed itinerary.

\section{The black-hole kink instantons}
\setcounter{equation}{0}

The Euclidean section of the Schwarzschild solution is asymptotically
flat and nonsingular because it does not contain any points
with $r<2M$. Thus, the curvature singularity does not lie
on the Euclidean section. Here I shall consider
the instantons that can be associated with the black hole kinks,
and show that their Euclidean sections can be extended beyond
the horizon down to the surface $r=A$.

The Euclidean continuation of the metrics which contain one kink
should be obtained by putting
\begin{equation}
\bar{t}=i\bar{\tau}.
\end{equation}
Using (2.6) and (2.7) we then have
\begin{equation}
d\bar{\tau}=-idT+\frac{ik}{\cos 2\alpha}dr,
\end{equation}
which is valid both for D=4 and D=5 black holes. This Euclidean
continuation would give rise to metrics which are positive
definite if we choose either the usual continuation $T=i\tau$,
for $r\geq 2M$, or the new Euclidean continuation $r=-i\rho$,
$M=-i\mu$, for $r<2M$, where $r$ becomes timelike, and we
transform a space coordinate into a time coordinate. In the
first case, restricting to D=4, metric (2.9) becomes
\begin{equation}
ds^{2}=\cos 2\alpha d\bar{\tau}^{2}\mp 2ikd\bar{\tau}dr+r^{2}d\Omega_{2}^{2}.
\end{equation}
This corresponds to the usual Euclidean subsection $\infty\geq r\geq2M$,
\begin{equation}
ds^{2}=\left(1-\frac{2M}{r}\right)d\tau^{2}
+\left(1-\frac{2M}{r}\right)^{-1}dr^{2}+r^{2}d\Omega_{2}^{2},
\end{equation}
and can be maximally-extended to the Kruskal metric
\begin{equation}
ds^{2}=-\frac{32M^{3}k}{r}e^{-\frac{r}{2M}}d\tilde{U}d\tilde{V}
+r^{2}d\Omega_{2}^{2},
\end{equation}
where
\begin{equation}
\tilde{U}=\mp e^{\frac{i\tau}{4kM}}e^{\frac{r}{2M}}\left(\frac{2M-r}{M}\right),
\;\;\; \tilde{V}=\mp\frac{k}{2}e^{-\frac{i\tau}{4kM}}.
\end{equation}

In order for the new continuation $r=-i\rho$, $M=-i\mu$ to give rise
to a metric which is positive definite for $r<2M$, we would also
continue the angular polar coordinates such that $\theta=-i\Theta$,
$\phi=\phi$. With this choice we then had for the orthogonal
coordinates
\[x=-X=-\rho\sinh\Theta\cos\phi, \;\;
y=-Y=-\rho\sinh\Theta\sin\phi,\]
\[z=-iZ=-i\rho\cosh\Theta,\]
so that $r=\sqrt{(x^{2}+y^{2}+z^{2})}=\pm i\rho$ and $\mid Z\mid\geq\rho$.
Therefore, we in fact have $\phi=\arctan\frac{y}{x}=\arctan\frac{Y}{X}$,
and $\theta=\arccos\frac{z}{r}=\pm i\Theta$,
with $\Theta=\cosh^{-1}\frac{Z}{\rho}$. Hence, the metric on the unit
two-sphere $d\Omega_{2}^{2}$ should transform as

\[d\Omega_{2}=\pm id\omega_{2}=i(d\Theta^{2}+\sinh^{2}\Theta d\phi^{2})^{\frac{1}{2}}.\]
The choice of the minus sign for the Euclidean continuation of both
the radial coordinate $r$ and the polar angle $\theta$ would allow us
to have the same action continuation as that corresponding to the
continuation $T=i\tau$; i.e. $S=iI$, where $S$ and $I$ are the
Lorentzian and Euclidean action, respectively, since the scalar
curvature transforms as $R(r)=-R(\rho)$ under continuation
$r=-i\rho$.

Thus, for the continuation $r=-i\rho$, $M=-i\mu$, $d\Omega_{2}
=\pm id\omega_{2}$ for $r<2M$, metric (2.9) becomes
\begin{equation}
ds^{2}=\cos 2\alpha d\bar{\tau}^{2}\pm 2kd\bar{\tau}d\rho+\rho^{2}d\omega_{2}^{2},
\end{equation}
which corresponds to the new Euclidean subsection $2M>r>M$, with
positive definite metric
\begin{equation}
ds^{2}=(\frac{2\mu}{\rho}-1)dT^{2}+(\frac{2\mu}{\rho}-1)^{-1}d\rho^{2}
+\rho^{2}d\omega_{2}^{2},
\end{equation}
and can be maximally-extended to the Kruskal metric
\begin{equation}
ds^{2}=+\frac{32\mu^{3}k}{\rho}e^{-\frac{\rho}{2\mu}}d\hat{U}d\hat{V}
+\rho^{2}d\omega_{2}^{2},
\end{equation}
where in this case
\begin{equation}
\hat{U}=\mp e^{\frac{\tau}{4k\mu}}e^{\frac{\rho}{2\mu}}\left(\frac{2\mu-\rho}{\mu}\right),
\;\;\; \hat{V}=\mp\frac{k}{2}e^{-\frac{\tau}{4k\mu}}.
\end{equation}

Using a positive definite metric such as (3.8) leads however
to the problem that the azimuthal angle $\theta$ is a periodic
variable only outside the horizon. In this case the transverse
two-manifold, which is a two-sphere of positive scalar curvature
outside the Euclidean horizon, becomes a hyperbolic plane of
negative scalar curvature inside the horizon and any boundary
at finite geodesic distance inside the horizon is no longer
compact. In particular, the boundary at $r=M$ would then have
the noncompact topology $S^{1}\times R^{2}$, with $S^{1}$
corresponding to time $T$. One cannot identify this geometry
at $\rho=2\mu$ with the geometry at $r=2M$ corresponding to
the compact topology $S^{1}\times S^{2}$ of the Gibbons-Hawking
instanton [18]. Nevertheless, avoidance of the spacetime
singularities in the calculation of the black hole action
does not actually require having a positive definite metric
in our spacetime kinks. Indeed, the "tachyonic continuation"
of the signature + + - - (which is - - + +) that
corresponds to a real azimuthal periodic variable $\theta$
also inside the horizon and implies the same metrics as
(3.7), (3.8) and (3.9) but with the sign for the polar
coordinate terms reversed, can not only avoid singularities
but erase them even at $r=0$. In order to see this, let us
consider the new variables $y+z=U$ and $y-z=V$ in the
Kruskal metric (2.21) which then becomes
\begin{equation}
ds^{2}=-\frac{32kM^{3}}{r}e^{-\frac{r}{2M}}(dy^{2}-dz^{2})+r^{2}d\Omega_{2}^{2},
\end{equation}
with
\begin{equation}
y^{2}-z^{2}=ke^{\frac{r}{2M}}\left(1-\frac{r}{2M}\right)
\end{equation}
\begin{equation}
\frac{y+z}{y-z}=ke^{\frac{\bar{t}}{2kM}}e^{\frac{r}{2M}}\left(\frac{r}{2M}-1\right).
\end{equation}
The singularity at $r=0$ lies on the surfaces $y^{2}-z^{2}=k$.
Although this singularity cannot be removed by any coordinate
changes, it can be avoided by defining either a new coordinate
$\zeta=iy$ or a new coordinate $\xi=iz$. For the first choice
the metric takes the Euclidean form
\begin{equation}
ds^{2}=\frac{32kM^{3}}{r}e^{-\frac{r}{2M}}(dz^{2}+d\zeta^{2})+r^{2}d\Omega_{2}^{2},
\end{equation}
which is positive definite in the patch $k=+1$ and 
has in fact signature
- - + + in the patch $k=-1$. The radial coordinate is then defined
by
\begin{equation}
z^{2}+\zeta^{2}=ke^{\frac{r}{2M}}\left(\frac{r}{2M}-1\right).
\end{equation}

On the section on which $z$ and $\zeta$ are both real (the usual
Euclidean section for patch $k=+1$) $\frac{r}{2M}$ will be real
and greater or equal to 1 on patch $k=+1$, and
$\frac{1}{2}\leq\frac{r}{2M}\leq 1$ on patch $k=-1$, the lower
limit $\frac{1}{2}$ being imposed by the continuity of the kink
at $\alpha=\frac{\pi}{2}$. Define the imaginary time by
$T=i\tau$. This continuation leaves invariant the form of
the metric (3.14) and is therefore compatible with the
coordinate transformation $\zeta=iy$. Then, from (2.20) and
(3.13) we obtain
\begin{equation}
z-i\zeta=\pm\left(z^{2}+\zeta^{2}\right)^{\frac{1}{2}}e^{\frac{i\tau}{4kM}}.
\end{equation}
It follows that for this time continuation $\tau$ is periodic
with period $8\pi kM$. On this nonsingular Euclidean section
(see Fig. 2a),
$\tau$ has then the character of an angular coordinate which
rotates about the "axis" $r=2M$ clockwise in patch $k=+1$,
and anticlockwise about the "axis" $r=0$ in patch $k=-1$.
Any boundary $\partial M_{k}$ in this Euclidean section has
topology $S^{1}\times S^{2}$ and so is compact in both
coordinate patches. Since the scalar curvature $R$ vanishes,
the action can be written only in terms of the surface
integrals corresponding to the fixed boundaries. This action
can be written
\begin{equation}
I_{k}=\frac{1}{8\pi}\int_{\partial M_{k}}d^{3}xK_{k},
\end{equation}
where $K_{k}=K-\frac{1}{2}(1+k)K^{0}$, $K$ being the trace
of the second fundamental form of the boundary, and $K^{0}$
the trace of the second fundamental form of the boundary
imbedded in flat space. This action was evaluated [16] in
the case of the positive definite metric which corresponds
to $k=+1$. It is $I_{+1}(M)=4\pi iM^{2}$. In the case $k=-1$,
fixing the boundary at the surface $r=A=M$, we also have
\[I_{-1}(M)=\frac{1}{8\pi}\int_{\partial M_{-1}^{A}}Kd\Sigma\]
\[=-4\pi i(2r-3M)\mid_{r=M}=4\pi iM^{2}.\]

For the second choice of coordinates, $\xi=iz$, metric (3.11)
takes the form
\begin{equation}
ds^{2}=-\frac{32kM^{3}}{r}e^{-\frac{r}{2M}}(dy^{2}+d\xi^{2})+r^{2}d\Omega_{2}^{2},
\end{equation}
which is positive definite in patch $k=-1$ and has again signature
- - + + in patch $k=+1$. The radial coordinate is now defined by
\begin{equation}
y^{2}+\xi^{2}=ke^{\frac{r}{2M}}\left(1-\frac{r}{2M}\right),
\end{equation}
so that on the section on which $y$ and $\xi$ are both real
(the usual Euclidean section for patch $k=-1$) $\frac{r}{2M}$
will be in the interval $\frac{1}{2}\leq\frac{r}{2M}\leq 1$
on patch $k=+1$, and greater or equal to 1 on patch $k=-1$.
We define now the imaginary $r$ and $M$ by
$r=-i\rho$ and $M=-i\mu$, keeping $T$ and the azimuthal coordinate
$\theta$ real. In order for this definition to be compatible with the
coordinate transformation $\xi=iz$, it should leave metric
(3.18) formally unchanged. For this to be accomplished one
must also continue the line element $ds$ itseft, namely
$ds=-id\sigma$, instead of the azimuthal angle $\theta$.
This requirement becomes most natural if we recall that the
interval $ds$ has the same physical dimension as that of $r$
and $M$, and that the "tachyonic" mass $\mu$ should be
associated with an imaginary relativistic interval. Then,
from (2.20) and (3.13) we obtain
\begin{equation}
y-i\xi=\pm(y^{2}+\xi^{2})^{\frac{1}{2}}e^{\frac{iT}{4k\mu}}.
\end{equation}
It is now the Lorentzian time $T$ which becomes periodic
with period $8\pi k\mu$. On this new nonsingular Euclidean
section (see Fig. 2b), 
$T$ would have the character of an angular coordinate
which rotates about the "axis" $\rho=0$ clockwise in the
patch $k=+1$, and anticlockwise about the "axis"
$\rho=2\mu$ in the patch $k=-1$. In such a new section, the
action is given by (3.17), where now
$K_{k}=K-\frac{1}{2}(1-k)K^{0}$. On the patch $k=+1$, we have
[18] 
\[I_{+1}(\mu)=\frac{1}{8\pi}\int_{\partial M_{+1}^{A}} Kd\Sigma \]
\[=4\pi iM(2r-3M)\mid_{r=M}=-4\pi iM^{2}=4\pi i\mu^{2}.\]
In the patch $k=-1$, taking $K^{0}=\frac{2}{r}$ and following
Gibbons and Hawking [18], we obtain the action $I_{-1}(\mu)$
which turns out to be the same as $I_{+1}(\mu)$.

Thus, on the coordinate patch $k=+1$, the Euclidean continuation
(3.1) of the time coordinate $\bar{t}$ of the kink metric
contains both the continuation for time $T$, $T=i\tau$,
where the apparent singularity at $r=2M$ is like the
irrelevant singularity at the origin of the polar coordinates
provided that $\frac{\tau}{4M}$ is regarded as an angular
variable and is identified with period $2\pi$ [18], and a new
continuation $r=-i\rho$, which also implies "tachyonic"
continuations $M=-i\mu$ and $ds=-id\sigma$, where the
curvature singularity at $\rho=0$ becomes again like a
harmless polar-coordinate singularity provided that
$\frac{T}{4\mu}$ is regarded as an angular variable and
is identified with period $2\pi$. The transverse two-manifold
is now a compact two-sphere both outside and inside the
Euclidean horizon and any boundaries have compact topology
$S^{1}\times S^{2}$, with $S^{1}$ corresponding to $\tau$
outside the horizon and to $T$ inside the horizon. Since
these topological products are compact, have the same
Euler characteristic and are both orientable, they are
homeomorphic to each other with a continuous mapping 
between them. Therefore, one can identify the two
corresponding geometries at the Euclidean horizons ($r=2M$
and $\rho=2\mu$) which, respectively, $\tau$ rotates about
at zero geodesic distance and is the geodesic distance at
which $T$ rotates about $\rho=0$. The Gibbons-Hawking
instanton can then be extended beyond the Euclidean
horizon down to just the boundary surface at $r=M$ ($\rho=\mu$)
where the first and second patches must be somehow joined
onto each other (Fig. 2c). 
The resulting Euclidean section does not
contain any points with $r<M$ and therefore the curvature
singularity is still avoided, as it also is in the baby
universe sector ($\rho < \mu$) due to 
the periodic nature of the instantonic
time $T$. The spacetime of the extended instanton covers
the entire domain of the coordinate patch $k=+1$ and that
of the baby universe is outside the two coordinate patches.

Euclidean continuation (3.1) on coordinate patch $k=-1$
leads to the same instantonic sections as for patch $k=+1$,
but now $T=i\tau$ corresponds to the section inside the
horizon $r=2M$ up to $r=M$, and $r=-i\rho$, $M=-i\mu$,
$ds=-id\sigma$ define the section outside the horizon
$\rho=2\mu$, with $\tau$ and $T$ respectively rotating about
$r=0$ and $2\mu$, anticlockwise in both cases. Since the
boundaries at constant radial coordinates on both sides
of the Euclidean horizon have compact topology $S^{1}\times
S^{2}$, the geometries at the Euclidean horizon ($r=2M$ and
$\rho=2\mu$) can also be identified, leading to an instanton
which covers the entire coordinate patch $k=-1$.

On any $\tau -r$ plane in the coordinate patch $k=+1$ we
can define the amplitude
$\langle\tau_{2}\mid\tau_{1}\rangle$ to go from the surface 
$\tau_{1}$ to the surface $\tau_{2}$ which is dominated by
the action $I_{1}(M)=4\pi iM^{2}$, corresponding to the
circular sector limited by the times $\tau_{1}$ and $\tau_{2}$
on a circle centered at $r=2M$ with large radius $r_{0}\gg 2M$.
Similarly, on the $t-\rho$ plane in the patch $k=+1$ the amplitude
$\langle t_{2}\mid t_{1}\rangle$ to go from the surface $t_{1}$
to the surface $t_{2}$ is dominated by the action $I_{2}(\mu)=4\pi
i\mu^{2}$ that corresponds to the sector limited by times
$t_{1}$ and $t_{2}$ from $\rho=\mu$ to $\rho=2\mu$ on a circle
centered at $\rho=0$. An asymptotic observer 
in patch $k=+1$ would interpret
these results as providing the probability of the occurrence
in the vacuum state of, respectively, a black hole with mass
$M$ or a white hole with mass $\mu$. In the coordinate patch
$k=-1$, the same observer would reach the same interpretation but
for a black hole with mass $\mu$ or a white hole with mass
$M$.

All of the above discussion can be readily extended to the
case of the hole kinks in five dimensions. We then have
similar instantons which contain both all points with
$r\geq M$ for $T=i\tau$ and all points with $\mu
>\rho\geq\frac{\mu}{\sqrt{2}}$ for $r=-i\rho$, $M=-i\mu$
and $ds=-id\sigma$ on the coordinate patch $k=+1$, and
all points with $\rho\geq\mu$ for $r=-i\rho$, $M=-i\mu$
and $ds=-id\sigma$ and all points with
$M>r\geq\frac{M}{\sqrt{2}}$ for $T=i\tau$ on the coordinate
patch $k=-1$.

Any constant time section of these instantons (Fig. 2) would
represent the half of respectively a three- and a four-dimensional
wormhole either in the first or the second coordinate patch.
The connection of one such wormhole halves in the first patch
to other wormhole half in the second patch would take place
on an equatorial surface $r=A$ and produce a complete
wormhole with two original asymptotic regions, one in patch
$k=+1$ and the other in patch $k=-1$. This connection will
be considered in more detail in Sec. V.

\section{\bf Quantum Theory of Black-Hole Kinks}
\setcounter{equation}{0}

\subsection{Conformal Structures and Foliations}

In Fig. 3 we show the Penrose diagram for the four-dimensional
black-hole kink, resulting from glueing at the identified
surfaces on $r=M$ the two coordinate patches, both on the
original and new, unphysical regions. The spacelike Cauchy
surfaces at constant time $T=t_{0}$ are labeled $\Sigma_{p/u}^{\pm}$,
depending on whether the original (physical, $p$) or new
(unphysical, $u$) region of the first (+) or second (-) patch
is being considered. These surfaces would correspond to
Einstein-Rosen bridges [14] with topology $R\times S^{2}$ in
each coordinate patch. A similar Penrose diagram can also be
constructed for the case D=5.

For Lorentzian four-dimensional
black-hole metric in the first coordinate patch
($t_{0}$ real), the Kruskal coordinates $U$, $V$ are both
pure imaginary for $r<2M$, and both real otherwise. Thus,
starting at spatial infinity of the physical region, the
$t_{0}$-surfaces follow their way first as $\Sigma_{p}^{+}$
or $\Sigma_{p}^{-}$, approaching the bifurcation point
$U=V=0$. At that point, before continuing through the
unphysical region, the surfaces can be analytically
continued into the upper ($k=+1$) or lower ($k=-1$)
imaginary plane $U^{*}$, $V^{*}$, where $U\rightarrow iU^{*}$,
$V\rightarrow iV^{*}$, $0\leq U^{*},V^{*}\leq 1$, as the
surface sector $\tilde{\Sigma}_{p}^{+}$ or $\tilde{\Sigma}_{p}^{-}$,
to finally reach the upper ($k=+1$) or lower ($k=-1$) surface
$r=A$; the surfaces then return to the bifurcation point
through the lower ($k=+1$) or upper ($k=-1$) imaginary
$U^{*},V^{*}$-plane as the sector $\tilde{\Sigma}_{u}^{+}$
or $\tilde{\sigma}_{u}^{-}$, and complete finally their
itinerary along the real unphyical region as $\Sigma_{u}^{+}$
or $\Sigma_{u}^{-}$ (Fig. 3). This itinerary would require
the identification of points on the upper surface at $r=A$
with the corresponding points (obtained by reflecting in
the bifurcation point $U=V=0$) on the lower surface also at
$r=A$, in each coordinate patch. It is worth noting that
these identifications give rise to the kind of periodicity that
would imply Hawking thermal effect [28] in each patch, which becomes
now a mathematical requirement of the present model. Thus, since
time $\bar{t}$ enters the Kruskal coordinates $U$, $V$ in the
form of the dimensionless exponent $\frac{\bar{t}}{4kM}$ for
D=4, one should generalize expression (2.20) to read
\begin{equation}
\bar{t}_{g}=\bar{t}+2\pi\kappa kiM(1-\kappa),
\end{equation}
where $\bar{t}$ is given by (2.20) and $\kappa=\pm 1$. For
$\kappa=-1$, the points $(\bar{t}-4\pi kiM,M,\theta,\phi)$
on each patch are in fact the points on the same patch, obtained
by reflecting in the bifurcation point, while keeping the
Kruskal metric real and unchanged.

On the extended instanton studied in the precedent section, each surface
\begin{equation}
\bar{\Sigma}^{\pm}=\Sigma_{p}^{\pm}\cup\tilde{\Sigma}_{p}^{\pm}
\cup\tilde{\Sigma}_{u}^{\pm}\cup\Sigma_{u}^{\pm}
\end{equation}
is spacelike everywhere in each coordinate patch, as their slope
does not change sign along the paths. Surfaces $\bar{\Sigma}^{\pm}$
allow therefore a complete foliation in each patch separately. Each
of the $\bar{\Sigma}^{+}$ represents a wormhole half (an Einstein-Rosen
bridge plus half an internal throat) in the patch $k=+1$, and each
of the $\bar{\Sigma}^{-}$ represents a similar wormhole half in the
patch $k=-1$. The discrete isometry $U\rightarrow -U$,
$V\rightarrow-V$ transforms $\bar{\Sigma}^{\pm}$ into itself,
so that asymptotically flat physical regions are mapped onto
asymptotically flat unphysical regions, and physical wormhole
necks are mapped onto unphysical wormhole necks.

The points at given constant time $T$ on the
minimal surfaces at $r=M$ of these wormhole halves in either
patch cannot be joined to the corresponding minimal surfaces at
the same constant time $\bar{t}_{0}$ in the other patch. Therefore,
although each coordinate patch admits a complete foliation by
spacelike surfaces of constant $T=t_{0}$ on the instanton, 
it is not possible to
foliate the entire spacetime corresponding to the complete
Penrose diagram formed by joining the diagrams for the two patches
with spacelike curves of constant $T$.

One still might try a foliation in terms of surfaces of constant
time $t$ which is defined so that
\begin{equation}
dT=dt+\tan 2\alpha dr.
\end{equation}
Then, by integrating (4.3) using the change of variable $p=k\cot\alpha$
and $\sin\alpha=\sqrt{\frac{M}{r}}$, we obtain
\[\bar{t}=t+4M\cot\alpha -Mk\csc^{2}\alpha\]
\[+4Mk\ln \left( \left| \left[ \frac{(\sin\alpha-k\cos\alpha)
\sin^{2}\alpha}{(\sin\alpha+k\cos\alpha)(2\sin^{2}\alpha -1)}
\right] ^{\frac{1}{2}} \right| \right) \]
\begin{equation}
+2Mk\kappa i(1-\kappa)\pi ,
\end{equation}
with $\kappa=\pm 1$ as in (4.1). 

In terms of time $t$, metric (2.4) is transformed into
metric (2.5)
which is not static and does not explicitly distinguish the two
coordinate patches.

There are Cauchy surfaces of constant $t$, both on the physical and
unphysical regions. These surfaces go from spatial infinity to the
surface $r=M$, after crossing the horizon at $r=2M$ (Fig. 3), 
but are not
spacelike everywhere as their slopes change sign at the event
horizon. However, once the surfaces have reached the horizon from
the physical region, they
can be continued first along such a horizon, passing over the
bifurcation point $U=V=0$, until they intersect the crossings of
the corresponding surfaces in the unphysical region with the
horizon, to go then along the latter surfaces into unphysical
spatial infinity. The resulting spacelike surfaces,
$\Sigma_{t}^{\pm}$, can therefore foliate the entire exterior
regions of both coordinate patches, but not their interior
regions.

A completely parallel treatment can be made starting with the
Lorentzian five-dimensional black hole, taking now $\bar{t}_{0}$
as a real quantity for $k=+1$. Also in this case the
spacetime can be foliated on the instanton
in each coordinate patch separately
by surfaces of $T=\bar{t}_{0}$ constant. Similarly, foliation
by surfaces of constant $t$, which enters a metric like (2.5)
for a three-sphere, instead of a two-sphere, is only possible
in the exterior regions of each coordinate patch, but not in
their interiors.

\subsection{Pure Quantum States}

Quantum states can be introduced only on those specetime regions
which are foliable by a family of spacelike surfaces [29]. However,
neither the Cauchy surfaces $\bar{\Sigma}^{\pm}$ nor the Cauchy
surfaces $\Sigma_{t}^{\pm}$ can foliate the full spacetime which
is covered by the coordinate patches when joined on the surface
at $r=A$, with $A=M$ for D=4 or $A=\frac{M}{\sqrt{2}}$ for D=5.
Then, there could not exist well-defined quantum states for the
whole of this spacetime construct, but only for each of the two
spacetime patches separately.

Let us consider those situations for which the Kruskal coordinates
$U$, $V$ are both imaginary for $A\leq r\leq r_{h}$, where either
$A=M$, $r_{h}=2M$, or $A=\frac{M}{\sqrt{2}}$, $r_{h}=M$, depending on
whether the D=4 case or the D=5 case is being considered. These
situations would correspond to $\bar{t}_{0}$ being complex with
imaginary part $2\pi M$ in the D=4 case, and to $\bar{t}_{0}$
being real in the case D=5. From the perspective of an observer
in each of the asymptotic regions, one should divide the full
spacetime at $r=A$ in two disconnected patches. Each of the
two resulting lone black holes, long after it is formed by an
independent collapse will possess states which are describable
by a static geometry and small perturbations. In addition to
one of such physical black holes, there will also be at late times
an unphysical black hole obtained by maximally extending the first
black hole spacetime. The maximally-extended solution for a
static black hole is what is known as an eternal black hole [30].
In this case, the data on the Cauchy surfaces $\Sigma_{u}^{\pm}$
cannot influence the black hole exterior and therefore any
perturbations would propagate to the future enterely inside the
horizon. Thus, the $\Sigma_{u}^{\pm}$ will lie in the Lorentzian
interior of the black hole. The boundaries $\Sigma^{\pm}=
\Sigma_{u}^{\pm}\cup\Sigma_{p}^{\pm}$ are Einstein-Rosen bridges.

If we considered Couchy surfaces $\Sigma^{\pm}$ with the topology
of just an Einstein-Rosen bridge, without any continuation into
the physical black-hole interior, then one could define quantum
states on surfaces $\Sigma^{\pm}$ which foliate just the regions
with $r\geq r_{h}$. Such states may be given as a pure-state
wave functional satisfying a no-boundary condition on the
event horizon, where we in fact have no singularity or edge.
This wave function would represent some ground state for the
black hole and be given in terms of the (D-1)-geometry and
regular matter fields on surfaces $\Sigma^{\pm}$ with
topology
$R\times S^{D-2}$. It therefore could be expressed as an
Euclidean path integral of Hartle and Hawking over D-geometries
and space-time matter fields bounded by $\Sigma^{\pm}$ and D-dimensional
asymptotically flat and empty infinity. Such a proposal has in
fact been already advanced by Barvinsky, Frolov and Zelnikov [31].

However, the analytic continuation of the kink time $\bar{t}$
allows one to extend the Euclidean section beyond $r=r_{h}$
up to $r=A$, so that the Gibbons-Hawking instanton would
correspondingly be extended to embrace all points with 
$r\geq A$. The half of this extended instanton is depicted
in Fig. 2c. Now the surface $\bar{\Sigma}^{\pm}$ becomes one
boundary of the Euclidean manifold, and spatial infinity
$\partial M_{\infty}^{\pm}$ plus a new boundary  
$\partial M_{A}^{\pm}$ is another. The new boundary
$\partial M_{A}^{\pm}$ has the topology $T\times S^{D-2}$,
with $S^{D-2}$ a (D-2)-sphere with constant radius $r=A$.
Consequently, extended
quantum states can also be introduced on the surfaces 
$\bar{\Sigma}^{\pm}$ that foliate the whole of the spacetime
corresponding to each coordinate patch separately. Such
quantum states should then be given by a pure-state wave
function $\Psi^{\pm}$ which, in addition to information
for $r\geq r_{h}$, would contain information on the
spacetime region between $r=r_{h}$ and $r=A$. This wave
function cannot satisfy a no-boundary condition as there
is a boundary $\partial M_{A}^{\pm}$ at $r=A$.

$\Psi^{\pm}$ would be a functional of the (D-1)-geometry and
matter fields on surfaces $\bar{\Sigma}^{\pm}$, given by an
Euclidean path integral over D-geometries and spacetime matter
fields bounded by these $\bar{\Sigma}^{\pm}$, D-dimensional
asymptotically flat and empty infinity and the boundary
at $r=A$, such that as the
(D-1)-geometries approach the surfaces at $r=A$ the wave
function should express the fact that the D-metric must be
non-singular there, i.e.: we can choose as the boundary
condition at the inner boundary the simple general requirement
that the wave function should be regular. Since there must be
no excitations in the asymptotically flat and empty infinity,
$\Psi^{\pm}$ would correspond to a vacuum wave function
expressible as an Euclidean path integral
\begin{equation}
\Psi^{\pm}=\int_{C^{\pm}}DgD\phi e^{-I},
\end{equation}
where the choice of the sign in the superscript of $\Psi$ and $C$
will depend on whether the state functional for the first (+)
or second (-) patch is being considered. This path integral is
over the class $C^{\pm}$ of D-geometries and matter fields
$\phi$ bounded by $\bar{\Sigma}^{\pm}$, the D-dimensional
asymptotically flat and empty infinity and the surface
at $r=A$, i.e.: we choose for
the total boundary of (4.5) the boundary $\partial M^{\pm}
=\bar{\Sigma}^{\pm}\cup\partial M_{\infty}^{\pm}\cup\partial
M_{A}^{\pm}$. In Eqn.
(4.5) $I$ is an Euclidean action which can be generically
written as

\[I=-\frac{1}{16\pi}\int
dT^{*}d^{D-1}X^{*}\sqrt{g^{*}}\left(R^{*}
-16\pi L^{*}(\phi)\right)\]
\begin{equation}
+\frac{1}{8\pi}\int d^{D-1}X^{*}\sqrt{h^{*}}K^{*},
\end{equation}
and is obtained from the Lorentzian action $S$ by the continuation
$S=iI$ implied by $\bar{t}=i\bar{\tau}$. From (4.6) one can compute
both the Euclidean action in the subsection $\infty>r\geq r_{h}$
for $T^{*}=\tau, X^{*}=x, R^{*}=R(r,M), L^{*}=L, K^{*}=-K, 
g^{*}=g(r,M)$ and $h^{*}=h(r,M)$, and the Euclidean action in
the other subsection $r_{h}>r\geq A$ for $T^{*}=T, X^{*}=-X,
R^{*}=-R(\rho,\mu), L^{*}=-L, K^{*}=-K, g^{*}=g(\rho,\mu)$ and
$h^{*}=h(\rho,\mu)$, with $L$ the Lagrangian for the matter
fields, $K$ the trace of the second fundamental form, and
$h$ the determinant of the first fundamental form. The Euclidean action 
as given by (4.6) would be divergent because it corresponds to
an asymptotically flat spacetime [32]. Therefore, we must supplement
it with an additional boundary term for $K^{*}_{0}$ defined
on an asymptotic (D-1)-surface embedded in the asymptotically
flat spacetime. This term transforms like the surface term in
(4.6) for the two Euclidean subsections and therefore renders
the full action finite [32].

Explicit computation of the path integral (4.5) would require
specifying the integration measure in it. This can be generally
achieved by computing the path integral as a sum of all contributions
coming from all asymptotically flat (D-1)-metrics which satisfy
the Fadeev-Popov gauge-fixing conditions which remove the gauge
freedom associated with invariance of the full action under
arbitrary change of spacetime coordinates [33]. We do not try this
calculation here. The wave functions $\Psi^{\pm}$ will be
obtained by a different procedure in the next subsection.

One can also introduce density matrices $\rho_{p/u}^{\pm}$
for the quantum state of a single black hole or wormhole
half on each patch as a path integral analogous to (4.5) on
the whole instanton, rather than half the instanton, by tracing
over the values of the matter field on either the entire
surfaces $\bar{\Sigma}_{u}^{\pm}$ or $\bar{\Sigma}_{p}^{\pm}$.
We have
\[\rho_{p/u}^{\pm}=Tr_{u/p}\vert\Psi^{\pm}\rangle\langle\Psi^{\pm}\vert \]
\begin{equation}
=\int D\phi_{p/u}\Psi_{p/u}^{\pm *}(\phi_{p/u},\phi_{u/p}')\Psi_{p/u}^{\pm}(\phi_{p/u},\phi_{u/p}).
\end{equation}
Like for the wave functional (4.5), the density matrices (4.7) would
represent a pure state on each coordinate patch.

\section{Neutral black-hole pairs}
\setcounter{equation}{0}

Let us analyse the possibility for the formation of neutral
black-hole pairs which is suggested by the continuity of
the angle of tilt of the light cones at $\alpha=\frac{\pi}{2}$
on the surfaces $r=A$, connecting a coordinate patch to another.
Such surfaces would thus be identified on the Kruskal diagrams
of Fig. 1,
both on the original regions and the new, unphysical regions
created by Kruskal extension. Each point on $r=A$ simultaneously
belonging to both coordinate patches represents a (D-2)-surface
defined at the same value of time $\bar{t}$ and different values
of times $T$ and $t$ in the two patches. These identifications
amount to the existence of bridges or throats with size $A$,
connecting the physical (unphysical) interior of a black hole in patch
$k=+1$ to the physical (unphysical) interior of another black hole
in patch $k=-1$. It will be argued later on in this section
that, from the perspective of an observer in either of the
asymptotic regions, $\partial M_{\infty}^{+}$ or
$\partial M_{\infty}^{-}$, if the topological charge of
the kink is positive, the whole spacetime of the above
geometrical construct can be regarded as a black-hole
pair formed up by a black hole with positive mass $M$ in
the coordinate patch of the observer, and an {\it anti-black
hole} with negative mass -$M$ in the patch where the observer
is not. If we look at the coordinates of each patch as
describing a single universe that contains at least one
black hole, then for asymptotic observers in either patch
it would follows that a neutral black-hole pair could
exist provided each black hole in the pair is in a different
universe, and that the Newtonian interaction between the
two holes of a four-dimensional pair 
would be accelerating these holes
away from each other with a maximum force
\[F=\frac{M^{2}}{\left[(\delta r)_{k=+1\rightarrow
k=-1}\right]^{2}} =
\frac{M^{2}}{\left[(\delta r)_{k=-1\rightarrow
k=+1}\right]^{2}} = \frac{1}{4},\]
where Eqn. (A.10) of the Appendix has been used.

To an observer in an asymptotic universe, any knowledge
about the existence and properties of a black hole in
another universe, or about the meaning of the relative
motion associated with force $F$ is classically forbidden,
since classically the two black holes cannot be brought
into the asymptotic universe of the observer. Such a
knowledge may still be allowed however if, as it is
actually the case, the compact region beyond the horizon
for the asymptotic observer, running classically from
$r=2M$ to the curvature singularity, becomes the support
of an one-kink because such a region, which 
is here considered to be essentially
quantum-mechanical, actually contains all of the information
which is classically forbidden, including the black hole
with negative mass. This hole can be thus quantum-mechanically
brought into the asymptotic universe of the observer.

Two neutral black holes are expected not to be pair created if
they are both in the same universe. Even spontaneous creation
of neutral black-hole pairs driven by 
the effective cosmological constant in
an inflationary universe should be expected to be highly
suppressed [8].

From the discussion in Sec. II
it follows that there could be no family
of spacelike surfaces of constant time $T$ able to foliate the whole of 
the spacetime of one such black-hole pairs. Since points on
surfaces $r=A$
simultaneously belonging to the two coordinate 
patches are labeled by equal values of time $\bar{t}$,
but not $T$ or $t$, (D-1)-wormholes defined on $T=$const.
surfaces $\bar{\Sigma}^{\pm}$ cannot be continued from
one coordinate patch to another at any surface, unless
topology change transitions, including nonsimply connected
surfaces that extend down to $r=A$ with topology more complicated
than that of surfaces $\bar{\Sigma}^{\pm}$, sharing in common
only the asymptotically flat behaviour at infinities, also
contribute the path integral. Consequently, cutting at
$r=A$ on any surface $\bar{\Sigma}=\bar{\Sigma}^{+}\cup\bar{\Sigma}^{-}$
would not actually divide the corresponding wormhole manifold in two
disconnected parts. Therefore, the instanton associated to
the creation or annihilation of neutral black-hole pairs
should necessarily contain contributions from all those
topologies, other than that of $\bar{\Sigma}$, allowing for smooth
transitions from one black hole in the pair to another; i.e.:
such an instanton would correspond to the topology
of 
$\Sigma=\partial M_{\infty}^{+}\cup\Sigma^{pair}\cup\partial M_{\infty}^{-}$,
with
\begin{equation}
\Sigma^{pair}=\bar{\Sigma}^{+}\cup
\left(\sum_{i}\sigma_{i}\right)\cup\bar{\Sigma}^{-},
\end{equation}
where the $\sigma_{i}$'s represent inner surfaces whose
topology, labeled $i$, is
different from that for a Einstein-Rosen bridge. Then, one
could not factorize the full quantum state of 
black-hole pairs or complete wormholes in a product of
pure-state wave functions [34]. In particular, for a black-hole pair
state $P$, we would generally have $P\neq\Psi^{+}.\Psi^{-}$.

\subsection{Thermal radiation}

In order to investigate further on the nature of these black-hole
pairs, let us at this point consider the process of thermal emission
of one such pairs by first rewritting the analytical expression
for $\bar{t}$ given by (4.4) for D=4 in a more convenient form.
We note that one would still recover (2.9) from the general
Kruskal metric [15]
\begin{equation}
ds^{2}=-2F(U,V)dUdV+r^{2}d\Omega_{2}^{2}
\end{equation}
if we re-define the Kruskal coordinates as follows:
\begin{equation}
U=\tilde{U}=\pm\kappa e^{\frac{\bar{t}_{c}}{4kM}}
e^{\frac{r}{2M}}(\frac{2M-r}{M})
\end{equation}
\begin{equation}
V=\tilde{V}=\pm k\kappa e^{-\frac{\bar{t}_{c}}{4kM}}
\end{equation}
where
\begin{equation}
\bar{t}_{c}=\bar{t}+4\pi iMk\kappa ,
\end{equation}
with $\kappa$ as defined in (4.1) and (4.4), and
$\bar{t}$ taken to be the real part of (4.4). This choice
leaves expressions for $UV=\tilde{U}\tilde{V}$, $F$, $r$ and
the Kruskal metrics (2.21) real and unchanged. For
$\kappa=-1$ Eqns. (2.19) become, respectively,
the sign-reversed to (5.3)
and (5.4); {\it i.e} the points $(\bar{t}-4\pi iMk_{1},r,\theta ,\phi)$
on the coordinate patches of Fig. 1 are the points on the new
regions $III_{k_{1}}$ or $IV_{k_{1}}$, on the same figure,
obtained by reflecting in the origins of the respective $U,V$
planes, while keeping metric (2.21)
and the time $t$ real and unchanged. This leads to
identifications of hyperbolae in the new, unphysical regions
$III_{k_{1}}$ and $IV_{k_{1}}$ with hyperbolae in,
respectively, original regions $II_{k_{1}}$ and $I_{k_{1}}$.

The evolution of a field along null geodesics as those in Fig.
1 can be described using a quantum propagator. If the field is
scalar with mass $m$, such a propagator will be the one used
by Hartle and Hawking [35] which satisfies the Klein-Gordon
equation
\begin{equation}
(\Box_{x}^{2}-m^{2})G(x',x)=-\delta(x,x').
\end{equation}
We note [35] that for metric (2.21) the propagator $G(x',x)$ will be
analytic on a strip of width $4\pi M$ which precisely is that
is predicted by the imaginary constant component of (5.5), thus
without any need of extending time $t$ into the Euclidean
region. Then, following Hartle and Hawking [35], the amplitude for
detection of a detector sensitive to particles of a given
energy $E$, in regions $I_{+}$ and $II_{-}$, would be
proportional to
\begin{equation}
\Pi_{E}=\int_{-\infty}^{+\infty}d\bar{t}_{c}e^{-iE\bar{t}_{c}}
G(0,\vec{R}';\bar{t}_{c},\vec{R}),
\end{equation}
where $\vec{R}'$ and $\vec{R}$ denote, respectively, 
$(r',\theta ',\phi ')$
and $(r,\theta ,\phi)$. Since time $\bar{t}_{c}$ (but not $t$)
already contains the imaginary constant term which is exactly
required for the thermal effects to appear, we need now not
make the time $t$ complex. From (5.5) and (5.7) one
can then write
\[\Pi_{E}=
e^{4\pi Mk_{1}k_{2}E}\]
\begin{equation}
\times\int_{-\infty}^{+\infty}d\bar{t}
e^{-iE\bar{t}}G(0,\vec{R}'; 
\bar{t}
+4\pi iMk_{1}k_{2},\vec{R}).
\end{equation}

Let us next consider a point $x'$ on the hyperbola $r=M$ of
region $II_{+}$, corresponding to patch $k=+1$. Since
such a hyperbola should be identified with the hyperbola
at $r=M$ of region $I_{-}$ of patch $k=-1$, the point
$x'$ can be taken to simultaneously belong to the two patches
at the same value of time $\bar{t}$.
Then, one can draw null geodesics starting at $x'$ which
connect such a point with different points $x$ on either
the original region $I_{+}$ of patch $k=+1$ or the original
region $II_{-}$ of patch $k=-1$. In the first case, we
obtain from (5.8)
\begin{equation}
P_{a}^{I_{+}}(E)=e^{-8\pi ME}P_{e}^{I_{+}}(E),
\end{equation}
where $P_{a}^{I_{+}}(E)$ denotes the probability for detector to
absorb a particle with positive energy $E$ from region $I_{+}$,
and $P_{e}^{I_{+}}(E)$ accounts for the similar probability
for detector to emit the same energy also to region $I_{+}$,
in the coordinate patch $k=+1$ corresponding to the
first universe. An observer in the exterior original region
of patch $k=+1$ will then measure an isotropic background
of thermal radiation with positive energy, at the Hawking
temperature $T_{BH}=(8\pi M)^{-1}$.

For the path connecting $x'$ with a point $x$ on the original
exterior region $II_{-}$ of patch $k=-1$ in the other
universe, we obtain for an observer in this region,
\begin{equation}
P_{a}^{II_{-}}(-E)=e^{+8\pi ME}P_{e}^{II_{-}}(-E).
\end{equation}
According to (5.10), in the exterior region $II_{-}$ of the
second universe there will appear as well an isotropic
background of thermal radiation, also at the Hawking
temperature $T_{BH}$, which is formed by exactly the
antiparticles to the particles of the thermal bath detected
in region $I_{+}$.
For $\kappa=+1$ we obtain similar hypersurface identifications
as for $\kappa=-1$. In this case, the identifications come
about in the situation resulting from simply exchanging the mutual
positions of the original regions $I_{k_{1}}$ and $II_{k_{1}}$
for, respectively, the new regions $III_{k_{1}}$ and
$IV_{k_{1}}$, on the coordinate patches of Fig. 1, while
keeping the sign of coordinates $U$, $V$ unchanged with
respect to those in (2.19); {\it i.e.} the points
$(\bar{t}+4\pi Mik_{1},r,\theta,\phi)$ on the so-modified
regions are the points on the original regions $I_{k_{1}}$
or $II_{k_{1}}$, on the same patches, again obtained by
reflecting in the origins of the respective $U$, $V$ planes,
while keeping metric (2.21) and the time $t$
real and unchanged. Thus, expressions for the relations between
probabilities of absorption and emission for $\kappa=+1$
are obtained by simply replacing region $I_{+}$ for
$IV_{+}$, region $II_{-}$ for $III_{-}$, and energy $E$
for $-E$ in (5.9) and (5.10), so that the same Hawking
temperature $T_{BH}$ is obtained in all the cases. Hence,
``observers'' will detect thermal radiation with energy $E<0$
on region $IV_{+}$, and with energy $E>0$ on region $III_{-}$,
in both cases at the Hawking temperature.

An evaporation process in patch $k=+1$ would then follow
according to which a black hole in a universe will disappear
completely, taking with it all the particles that fell in
to form the black hole and the antiparticles to emit
radiation, by going off through the wormhole of throat
radius $M$ whose other end, which opens up in other
universe described in the patch $k=-1$, 
is another black hole which can be regarded to
have been formed in patch $k=-1$
from the collapse of massive antifermions,
and also evaporated, giving off exactly the antiparticles
to the particles of the thermal radiation emitted by the
first black hole in the first universe [9]. 
One can then represent neutral black-hole pairs as instantons
with zero total energy. An asymptotic observer in one universe
would {\it interpret} each such pairs as being formed up by one
black hole with positive mass $M$ in the same universe, and
one anti-black hole with the corresponding negative mass
-$M$ in a different universe, both holes being connected
to each other by a wormhole with size $M$, and accelerated away
from each other with maximum force $F=\frac{1}{4}$. What
such an observer would actually observe out from the pair
is only either a black hole with positive energy or a
wormhole mouth opening to the observer's universe.

Interpreting that the black hole in the "other universe" has
negative mass without violating the well-established 
positive-energy theorems [36] can still be consistently
accomplished as follows. Let $({\bf M},g_{ab})$ be a
D-dimensional asymptotically flat spacetime and let
$\Sigma\subset {\bf M}$ be a spacelike hypersurface with
the pair $(\Sigma ,g)$ describing a kink with topological
charge $kink(\Sigma ;g)=\pm 1$, such that $(\Sigma
,h_{ab},K_{ab})$ is an asymptotically flat initial data
set for the first coordinate patch only, where $h_{ab}$
and $K_{ab}$ are the D-1 metric and the second fundamental
form on the hypersurface $\Sigma$. Then, one can generally
define the classical ADM mass (total mass) by [37]
\begin{equation}
E=\frac{1}{16\pi}\lim_{r\rightarrow\infty}\sum_{\nu=1}^{D-1}
\int\left(\frac{\partial h_{\mu\nu}}{\partial x^{\mu}}
-\frac{\partial h_{\nu\nu}}{\partial x^{\nu}}\right)N^{\nu}dA,
\end{equation}
in which $r$ is the radial coordinate, the integral is
taken over a sphere of constant radius $r$, and $N^{\nu}$
is the unit outward normal to this sphere. Choosing $\Sigma$
to be asymptotically orthogonal to the timelike Killing
field makes (5.11) agree with the Komar definition of mass [38]
as the total {\it outward} force that must be exerted by
the asymptotic observer to keep in place a mass density
distributed on the given spherical surface.

To the asymptotic observer in one coordinate patch it is
then clear that any mass distributed on surfaces everywhere 
in the same patch should be defined by (5.11) because,
respect to that observer, time must flow toward the future
for $\alpha\leq\frac{\pi}{2}$, so any mass distribution
satisfies the positive-mass theorems. However, for spherical
distributions of mass in the other patch, the same observer
would interpret ADM mass to be the sign-reversed to the
one given by (5.11): since for such an observer time at
any surface in the other patch of the kink
flows toward the past, he would interpret any ADM mass in
the other patch as a measure of the total {\it inward} force
that he had to exert on the mass distribution to keep it
in place. Clearly, to an asymptotic observer in the other
patch, what had negative mass to the first observer would
become of positive mass, and vice versa. Any asymptotic
observer in either patch would therefore attribute a
conserved vanishing mass to either a neutral black- or
white-hole pair, or a complete wormhole or a baby universe
living outside the spacetime of the kink.

\subsection{Mixed quantum states}

Let us return to the definition of the
black-hole pair quantum state. Since cutting at $r=A$ on any
surface $\bar{\Sigma}=\bar{\Sigma}^{+}\cup\bar{\Sigma}^{-}$
would not divide the corresponding (D-1)-wormhole manifold
into two parts, for the state $P$ of a black-hole pair one
should take a {\it nonfactorizable} density matrix that
describes a statistical mixed state. This can be given by
an Euclidean path integral [34]
\begin{equation}
P=\int_{C}DgD\phi e^{-I},
\end{equation}
over the class $C$ of D-geometries and matter field
configurations $\phi$ which are bounded by the prescribed
(D-1)-data on $\bar{\Sigma}^{+}\cup\partial M_{A}^{+}$ and
the D-dimensional asymptotically flat and empty infinity
$\partial M_{\infty}^{+}$,
and the orientation
reverse of the (D-1)-data on $\bar{\Sigma}^{-}\cup\partial
M_{A}^{-}$, and
the D-dimensional infinity $\partial M_{\infty}^{-}$,
$\bar{\Sigma}^{+}\cup\partial M_{A}^{+}$ and
$\bar{\Sigma}^{-}\cup\partial M_{A}^{-}$ together forming the
inner boundary $\bar{\Sigma}\cup\partial M_{A}$, 
and $\partial M_{\infty}^{+}$
and $\partial M_{\infty}^{-}$ together forming the asymptotic
boundary $\partial M_{\infty}$; i.e.: we choose as total
boundary of (5.12) $\partial M=\bar{\Sigma}\cup\partial
M_{\infty}\cup\partial M_{A}=\partial M^{+}\cup\partial M^{-}$. 
This nonfactorizable, mixed density matrix can
be obtained as the real part of the propagator, $K(+,-)$,
from the wave function $\Psi^{+}$ on
$\partial M^{+}$
at Euclidean time $\tau_{k=+1}$ to the wave function $\Psi^{-}$
on $\partial M^{-}$ 
at Euclidean time $\tau_{k=-1}$.
Assuming that a black hole possesses a discrete energy spectrum
labeled by the index $n$, this propagator can be written in
the form
\begin{equation}
K(+,-)=\sum_{n}\Psi_{n}^{+}\Psi_{n}^{-}e^{-E_{n}\Delta\tau},
\end{equation}
where $\Delta\tau$ is the time separation between the
two boundaries $\partial M^{+}$ 
and $\partial M^{-}$, and
$E_{n}$ denotes eigenenergy of the assumed discrete spectrum
of the black hole. Now, since the two boundaries may have any time
separation $\Delta\tau$, in order to obtain the density matrix
$P$ one has to integrate propagator (5.13) over all possible
values of $\Delta\tau$. We restrict in what follows to the
four-dimensional case. Then, according to our discussion in
Secs. II and III, and corresponding to the Euclidean
continuation of time $\bar{t}$ in (2.20) for the extended
instanton of Fig. 2c, the time separation $\Delta\tau$ turns
out to be always complex and given by
\begin{equation}
\Delta\tau=\tau_{1}-iT_{1}
\end{equation}
when $T_{1}$ is expressed in terms of $M$ rather than $\mu$.
Since the exponent in (5.13) is then real in $\tau_{1}$
and imaginary in $T_{1}$, one should allow $\tau_{1}$ to
vary from 0 to $\infty$ and $T_{1}$ to vary along an
one-fourth of its period; i.e.: from 0 to $2\pi M$.
Inserting (5.14) into (5.13) and integrating over $\Delta\tau$
with these integration limits, we obtain for the real part
of the resulting expression,
\begin{equation}
P\propto\sum_{n}\frac{\Psi_{n}^{+}\Psi_{n}^{-}}{E_{n}}.
\end{equation}
Eqn. (5.15) can be consistently interpreted by considering that
the black holes in the pair are in the quantum state specified
by the wave function $\Psi_{n}$ with a relative probability
\begin{equation}
P_{n}\propto\frac{1}{E_{n}}.
\end{equation}
This relative probability is or is not positive definite and
convergent depending on the values of the eigenenergies $E_{n}$
allowed by the black-hole spectrum. Since time $\tau$ is
periodic with period $8\pi M$, choosing any arbitrary origin
for time $\tau$, common in the two patches, one can always write
$\tau_{1}=8\pi MN+\delta\tau$, where $N$ is any positive
integer and $\delta\tau$ any time difference between 0 and
$8\pi M$ which, in turn, should induce integration of (5.13)
between 0 and $8\pi M$, i.e. along one new complete time
period. $\tau_{1}=\infty$ would therefore correspond to
repeating an infinite number of times the full period
$8\pi M$.
Then, periodicity of $\tau$ should imply
that the final physical situations one would achieve 
either by going
along all the equivalent complete complex integration paths
$\Delta\tau$, or after completing the integration path just
along the imaginary part, $T_{1}$, of such paths, must be identified.
Integration of (5.13) over $T_{1}$ from 0 to $2\pi M$ 
for $\tau_{1}=0$ yields a density matrix
\begin{equation}
P_{R}=\sum_{n}\frac{\Psi_{n}^{+}\Psi_{n}^{-}\sin^{2}\left(\pi
ME_{n}\right)}{E_{n}}.
\end{equation}
Equalizing then (5.15) and (5.17) we obtain for the black-hole
spectrum
\begin{equation}
E_{n}^{\pm}=\pm\frac{1}{M}\left(n+\frac{1}{2}\right),
\end{equation}
i.e. the black holes behave like though they were harmonic
oscillators with positive or negative frequency
$\omega^{\pm}=\pm\frac{1}{M}$
where, according to the discussion on black-hole thermal
emission above, one must choose the sign depending on whether
the first (+) or second (-) patch is being considered, i.e.
\begin{equation}
\omega^{k}=\frac{k}{M}. 
\end{equation}
The energy associated to these
frequencies can be taken as an estimate of 
the minimum possible energy inside the black
hole which is compatible with the 
quantum relation between energy and wavelength [39]. It
is worth noticing that if one defines the angular frequency
$\Omega^{k}=\frac{\omega^{k}}{2\pi}$ and disregards zero-point
energy, the number of quantum states in a black-hole pair,
\[\sl N =\frac{M}{\Omega^{+}}-\frac{M}{\Omega^{-}}=4\pi
M^{2}, \]
becomes the Bekenstein-Hawking entropy formula.
A procedure to compute
the Euclidean action for a neutral black-hole pair is
given in the Appendix. Following this procedure we
obtain the Bekenstein-Hawking entropy in a more rigurous
fashion, and can interpret it as a count of the number of
states residing in the interior region between the event
horizon and the surface $r=M$ for a Schwarzschild black-hole
pair.

The wave function for single four-dimensional
black holes in a pair can then be written
\begin{equation}
\Psi_{n}^{\pm}=\Psi_{n}^{k}(R_{k})=(\pi)^{-\frac{1}{4}}
\frac{e^{-\frac{1}{2}R_{k}^{2}}}{2^{\frac{n}{2}}\sqrt{n!}} 
H_{n}(R_{k}),
\end{equation}
where the $H_{n}$'s are the Hermite polynomials and 
$R_{k}$ denotes
some radial coordinate defined in the first ($k=+1$) or second
($k=-1$) patch. The quantum states (5.20) are formally the
same as those obtained for the scale factor of a Tolman-Hawking
wormhole [10].
The density matrix for neutral four-dimensional
black-hole pairs or complete
wormholes can finally be written
\[P\equiv P(R_{+},R_{-})\]
\begin{equation}
=\frac{2kMe^{-\frac{R_{+}^{2}+R_{-}^{2}}{2}}}{\pi^{\frac{1}{2}}}
\sum_{n}\frac{H_{n}(R_{+})H_{n}(R_{-})}{2^{\frac{n}{2}}\sqrt{n!}(2n+1)}.
\end{equation}
Thus, although this density matrix cannot diverge, it still is
not positive definite and therefore transitions between the two
black holes in a neutral pair are nonunitary.

\subsection{Generalized Quantum Theory}

It was pointed out in section 3.1 that the external regions,
$r\geq r_{h}$, can be foliated by surfaces $\Sigma^{\pm}$
of constant $T$, or by surfaces $\Sigma_{t}^{\pm}$ 
with $t=$const. in both
coordinate patches separately.
Yet, these two foliating regions are mutually
separated by the interior
black hole regions 
where no foliation by spacelike surfaces
in terms of sections of constant time $T$ or time $t$
is possible in the Lorentzian regime.
Since for the conventional quantum mechanics
to be formulated in terms of the usual unitary evolution and
reduction of the state vector, the whole of the spacetime must
be foliable by spacelike surfaces [29], such a conventional quantum
theory is still unable to describe overall evolution in 
neutral black hole
pairs. Then, one would either recourse to the nonunitary
Euclidean density-matrix formalism of Sec. VB, or
use a generalized approach to quantum theory [40]. We
shall consider here the latter approach.

The generalized quantum theory we shall use consists of:
(1) a set of fine-grained histories which allow for the most
refined possible description of the system. For this set we take
sequences of sets of one-dimensional projections $Q$ on every
member of the foliating families of spacelike surfaces in
the exterior regions of patches $k=+1$ and $k=-1$. (2) Allowed
sets of coarse-grained alternative histories which generally
are partitions of the fine-grained sets into mutually exclusive
classes $\{C_{\alpha}\}$. Here, simple examples [40] of such
coarse-grained sets in a Heisenberg-like picture can be chains
of projections $Q$ in both patches for spacelike surfaces
$\sigma^{k}$
\begin{equation}
C_{\alpha}^{k}=Q_{\alpha_{\iota}}^{\iota}(\sigma_{\iota}^{k}) ...
Q_{\alpha_{1}}^{1}(\sigma_{1}^{k}),\;\;\;\;  \sigma_{j}^{+/-}>/<\sigma_{0}^{+/-},
\end{equation}
where the sign or relation in
$./.$ depends on whether the first or second
patch is being considered, the $\sigma_{0}$ are spacelike
surfaces just before the interior regions, the $\alpha$'s denote
[40]
specifications of the particular alternative in the set (``yes'' or
``no''),
and $\iota$ expresses that
there may be different sets at the different times $T$ or $t$ labeling
the spacelike surfaces $\sigma^{k}$. The projections satisfy [40]
\begin{equation}
Q_{\alpha_{\iota}}^{\iota}Q_{\beta_{\iota}}^{\iota}
=\delta_{\alpha_{\iota}\beta_{\iota}}Q_{\alpha_{\iota}},\;\;\;\;
\sum_{\alpha_{\iota}}Q_{\alpha_{\iota}}^{\iota}=I .
\end{equation}
The set of all possible sequences gives a set of alternative
histories $\{C_{\alpha}^{k}\}$ which satisfy
\begin{equation}
\sum_{\alpha}C_{\alpha}^{k}=I.
\end{equation}
The projections $Q_{\alpha_{\iota}}^{k}$ should evolve unitarily
according to conventional quantum theory in the exterior regions,
but do not in the interior regions.
(3) Since we want to preserve causality,
one would allow for a re-definition of
the final Hilbert space for the whole system compatible with
unitarity. The evolution through the internal regions should then
be given by a nonunitary operator $X$ which (i) is derived
from a transition matrix given by a sum-over-histories [40]
\[\langle\phi^{-}(x),\sigma^{-}\mid\phi^{+}(x),\sigma^{+}\rangle\]
\begin{equation}
=\int_{[\phi^{+},\phi^{-}]}d\phi\exp\left( iS[\phi(x)]\right),
\end{equation}
where $\phi^{\pm}\equiv\phi^{k}$ denotes a definite spatial
field configuration on the spacelike surface
$\sigma^{\pm}\equiv\sigma^{k}$, and $S$ is the Lorentzian action
functional for the given matter field, (ii) connects alternatives
in the exterior regions of patch $k=+1$ to those in the
exterior regions of patch $k=-1$ in such a way that a whole
history is $C_{\beta}^{\mp}XC_{\alpha}^{\pm}$, and (iii) is
invertible in order to account for the periodic character of
our fixed spacetime (This periodicity arises from continuity of
the angle of tilt at $\alpha=\frac{\pi}{2}$ which implies
that surfaces with $r=A$ in the two patches should be
identified both on the original and new regions,
see Fig. 3). Then, the decoherence
functional measuring interference between pairs of histories
in a coarse-grained set would be given by [41]
\[D(\beta ',\alpha ';
\beta,\alpha)\]
\begin{equation}
=Tr\left[X^{-1}C_{\beta '}^{-}
XC_{\alpha '}^{+}
\rho C_{\alpha}^{+\dagger}
X^{\dagger}
C_{\beta}^{-\dagger}
(X^{\dagger})^{-1}\right],
\end{equation}
which is complex, hermitian, positive and normalized. The functional
(5.26), moreover,
is linear in the initial density matrix $\rho$ which incorporates
the boundary condition for the whole system. Expliciting
the form of this density matrix would fix the form of the
decoherence functional.

We will take for the quantum state describing the $k=+1$
exterior regions
a functional of the three-geometry and matter field configurations
on a boundary surface
$\sigma^{+}\equiv\Sigma_{0}^{+}=\Sigma_{0p}^{+}\bigcup\Sigma_{0u}^{+}$,
where the $\Sigma_{0}^{+}$'s are surfaces for a given constant
time $T$ labeled 0, or a constant time $t$ which is also labeled 0.
This can be given as a pure-state wave function $\Phi_{0}^{+}$ defined
by the path integral over the physical degrees of freedom on the
foliating surfaces on the patch $k=+1$. 
We shall choose a {\it no boundary} quantum state
\begin{equation}
\Phi_{0}^{+}=\int_{C^{+}}DgD\phi e^{-iS\left[g_{\mu\nu},\phi\right]},
\end{equation}
which is over the class $C^{+}$ of four-geometries and regular
matter fields bounded by
$\partial_{+}M=\Sigma_{0}^{+}\bigcup\partial M_{\infty}^{+}$, i.e.
by any spacelike surface $\Sigma_{0}^{+}$ and the four-dimensional
asymptotically flat and empty infinity. 
Thus, a density matrix which
incorporates a no boundary condition on spacelike surfaces
in only the universe $k=+1$
can be constructed from $\Phi_{0}^{+}$ by tracing over
the values of the fields on the unphysical regions,
$\phi_{u\infty}^{+}$, relative to the single universe
\[\rho=Tr_{u}\mid\Phi_{0}^{+}\rangle\langle\Phi_{0}^{+}\mid\]
\begin{equation}
\equiv\int D\phi_{p}\Phi_{0}^{+\ast}(\phi_{p},\phi_{u}')\Phi_{0}^{+}(\phi_{p},\phi_{u}),
\end{equation}
which gives the
kernel of the propagation amplitude from the initial
state $\phi_{p}$ on the right asymptotic infinity to $\phi_{u}$
on the left asymptotic infinity, in Lorentzian space.

\section{Cosmological Implications}

The most popular boundary condition for the universe was
suggested by Hawking in 1981 [42]. It can be stated by saying
that the boundary condition of the universe is that it has
no boundary, and so expressed the view that the single
universe should be self-contained. One can regard the no
boundary proposal as the cosmological version of the
feature that the Euclidean section of the Schwarzschild
solution is nonsingular because it does not contain any
points with $r<2M$ and shows no boundary at the event
horizon. Beyond the horizon, black holes are thought to
be connected through wormholes to other universes. The
cosmological version of the Gibbons-Hawking instanton
ignores any aspects altogether of the existence of such
a connection, and therefore it entails the self-containedness
of the universe.

There exists a more direct analogy between the cosmological
no boundary proposal and the complete Euclidean section
of the black hole interior discussed in Sec. III.
Restricting to the kinkless case, where we do not shop-off
inside the horizon along the spacelike hypersurface of
constant radius $r=A$, but allow completion of the 
Euclidean section  up to $r=0$ in just the coordinate
patch $k=+1$ (see Fig. 2b), and continuing in $r$, $M$
and $ds$ with the coordinate choice $\xi=iz$ as in
Sec. III, we in fact obtain a black hole interior without
any boundary or singularity, even in the Lorentzian sector.
The cosmological version of this kinkless black-hole
interior would then correspond to a self-contained
universe as far as only one coordinate patch is required
to describe it.

However, restricting to the kinkless black-hole interior
analogy, and hence to cosmological scenarios which are
self-contained, implies the lossing of at least some of
the quantum richness arising in the kinky framework. We
will therefore consider the more general boundary
conditions which may be suggested from the analogy with
the complete black-hole one-kink.
The Euclidean continuation of the D=4 black-hole kink metric
leads to an instanton which contains no points with
$r<M$, rather than $r<2M$. At $r=M$ there exists no singularity
but there is a boundary connecting the universe where the hole
has emerged to other universes. Since this boundary does not
divide the whole manifold in that of a single universe and
the rest manifold, the universe can therefore no
longer be self-contained. This suggests 
a different alternate outcome to
the problem of the boundary conditions of the single universe.
It is that the boundary condition of the universe is that 
there is no boundary other than that for the throat of a
wormhole, i.e. the boundary $\partial M_{A}^{\pm}$.
In this approach the notion that there is just
one single universe should be replaced by 
a new notion which could be regarded as 
a necessary previous step in the way to the modern quantum
cosmological view that there exists one single {\it
metauniverse} [43], that is the set of all possible incipient,
inflating and large universes
allowed by quantum cosmology. The proposed
boundary condition would simply express the feature that although
the whole of the metauniverse may well be self-contained, any
single universe in it should not.

The no boundary wave function of Barvinsky, Frolov and Zelnikov [31]
for the ground state of black holes corresponds to a modified
version of the no boundary ground-state wave functional for
asymptotically flat three-geometries proposed by Hartle [44].
Similarly, the state (3.6) would correspond to a modified
version of the wave function for Euclidean quantum wormholes.
Thus, parallelly to as the cosmological version of the
Barvinsky-Frolov-Zelnikov or Hartle asymptotically flat wave
functions are not but the familiar Hartle-Hawking prescription [45]
for the quantum state of a compact universe, the cosmological
version of (3.6) would correspond to a picture in which the
path integral is over Euclidean four-geometries and matter
field configurations on spacetimes which are no longer compact,
but possess a boundary $\partial M_{A}^{+}$
that connects the single universe to
the other objects that may exist in 
a unique, self-contained metauniverse.
Once again, one would
emphasize that this is simply a proposal which cannot be deduced
from some other principle, but just suggested by the physics
of black hole kinks.

\section{Conclusions}

As a previous step to the quantization of black holes, this
paper examines the spacetime geometry and associated
causal structure of black-hole kinks in four and five
dimensions by enforcing the complete domain of such spacetimes
to be described in terms of Finkelstein-McCollum standard
kink metrics. Two coordinate patches are then required to
represent a complete one-kink in the studied cases. These
patches are joined at a spacelike hypersurface which is an
interior common surface of two black holes, each of which
lies on a different coordinate patch, without violating
energy-momentum conservation or inducing mass-inflation
processes that may destroy the causal structure 
associated with the considered geometry. We call the
resulting construct a neutral black-hole pair.

Nonsingular instantons for each of the black holes in
a kinky pair have been constructed both outside and
inside the event horizon, and purely imaginary finite
actions have been evaluated on the respective
complexified sections. We interpret these actions as
a measure of the probability of finding the corresponding
kink metrics in the vacuum state.

The quantum state of the black holes in one such pairs
has been formulated by using the Euclidean path integral
approach, with a multiply connected boundary at the interior
spacelike surface where the two coordinate patches are
identified. By evaluating the independent thermal-radiation
properties of the two black holes in the pair with respect
to an asymptotic observer in one patch, it is shown that the
black hole in the other patch behaves like though if it were
formed only by negative energy in an amount whose absolute
value equals that of the positive-energy black hole which
always is in the same patch as the observer. This is
regarded to be a consequence from the essential quantum nature
of the pair in relation with the rotation of the light cones
in the one-kink, and circunvents therefore any violation
of the classical positive-energy theorems.

The quantum state that corresponds to either a black-hole
pair or, equivalently, a complete wormhole has been also
formulated using the Euclidean approach. This state 
turns out to be a nonfactorizable, mixed density matrix,
rather than a pure-state wave functional, expressible
as a propagator for the single black-hole wave function
from some hypersurface in one patch to another hypersurface
in the other patch. The use of the periodic properties of
Euclidean time enabled us to show that the quantum state
of each black hole in the pair must be given in terms
of harmonic-oscillator wave functions for a fundamental
frequency which is positive if the black hole is in the
same coordinate patch as the asymptotic observer, and
negative otherwise. Since the spacetime of a black-hole
pair admits no complete foliation, in formulating the
quantum theory we have also employed a generalized 
formalism in terms of a decoherence functional.

Finally, we have discussed some possible cosmological
implications of our black-hole kink model, suggesting
that the boundary condition for a single universe which
is not self-contained is that there is no boundary other
than those corresponding to the cross sections of
wormholes.

\acknowledgements

\noindent 
This work was supported by DGICYT 
under Research Project N\mbox{$^{\underline{o}}$} PB94-0107-A.

\appendix
\section{Euclidean action of the Schwarzschild pair}
\setcounter{equation}{0}
\alfabet

It appears
interesting to calculate an exact expression for the
Euclidean action of a Schwarzschild pair. The issue would
relate with the problem of the origin of the black hole
entropy. Pair production appears to be
independent of Planck scale physics and therefore it should
be an unambiguous consequence from quantum gravity [7].

The amplitude for production of nonrotating, chargeless
black hole pairs can be calculated in the semiclassical
approximation from the corresponding instanton action
which is
\begin{equation}
S=-\frac{1}{16\pi}\int d^{4}x\sqrt{g}R
-\frac{1}{8\pi}\int d^{3}x\sqrt{h}K,
\end{equation}
where $K$ denotes the trace of the second fundamental form,
and $h$ is the metric on the boundary. We wish to find the
contribution of the two black holes
to action (A.1). This contribution will correspond
to substracting from (A.1) the flat metrics that the
instanton asymptotically approaches.

Since variation of the tilt angle $\alpha$
must be continuous at $r=M$,
the above contribution can be evaluated from the
corresponding enforced variation
in the time parameter of the instanton at $r=M$
that leads to a net variation of radial coordinate $(\delta r)$
also on $r=M$. This variation would be computed with respect
to a given coordinate patch, as one passes from that patch to
the other, and leaves unchanged the flat metrics. It would
correspond to a variation of action (A.1), $\delta S$, which
should be finite when evaluated at $r=M$,. 
In what follows we shall interpret
$\delta S|_{r=M}$ as the contribution of the two black holes
to the instanton action (A.1).

Let us then evaluate $\delta S$. Using Einstein equations we find
\[\delta S=-\frac{1}{8\pi}\int d^{3}x\delta(\sqrt{h}h^{ij}K_{ij})\]
\begin{equation}
=-\frac{1}{8\pi}\int d^{3}x\left( \sqrt{h}(\delta h^{ij})(K_{ij}
-\frac{1}{2}h_{ij}K)+\sqrt{h}h^{ij}\delta K_{ij} \right).
\end{equation}
The last term in the rhs of (A.2) can be written
\[h^{ij}\delta K_{ij}=\frac{1}{2}h^{ij}\frac{\partial}{\partial t}\delta h_{ij}
=\frac{1}{2}\frac{\partial}{\partial x^{l}}h^{ij}\dot{x}^{l}\delta h_{ij}
=\frac{1}{2}\frac{\partial A^{l}}{\partial x^{l}},\]
with $A^{l}=h^{ij}\dot{x}^{l}\delta h_{ij}$ a contravariant vector,
and the overhead dot meaning time derivative. Hence
\[h^{ij}\delta K_{ij}=\frac{1}{2}\frac{1}{\sqrt{h}}\frac{\partial}{\partial x^{l}}(\sqrt{h}A^{l}), \]
so that
\[\int d^{3}x\sqrt{h}h^{ij}\delta K_{ij}
=\frac{1}{2}\int d^{3}x\frac{\partial}{\partial x^{l}}(\sqrt{h}A^{l}).\]
This can be now converted into an integral over $A^{l}$ extended to
the 2-surface surrounding the boundary. Since variations of the field
are all zero on the boundary, this term must vanish and we have
finally
\begin{equation}
\delta S=-\frac{1}{16\pi}\int d^{3}x\sqrt{h}
(\delta h^{ij})(2K_{ij}-h_{ik}K).
\end{equation}

For the Schwarzschild metric, we then have
\[\delta S_{1}=\eta\frac{M}{\pi}\left[-\frac{1}{24}\rho^{-\frac{3}{2}}
+\rho^{-\frac{1}{2}}+\arctan (\rho^{\frac{1}{2}})\right]\times\] 
\begin{equation}
\int_{0}^{\pi}d\theta\int_{0}^{2\pi}d\phi (\delta r),
\end{equation}
where $\rho =\frac{2M}{r}-1$ and $\eta=1$ for $r<2M$ and $\eta=i$
for $r>2M$. However, this variation of the action is still
divergent on the asymptotic boundary, and one should
supplement it with an additional variation, $\delta S_{0}$,
which corresponds to those terms in (A.4) that diverge at
infinity for finite $\delta r$. It can be seen that
$\delta S_{0}=\frac{\eta M}{\pi}\arctan(\rho^{\frac{1}{2}})$,
so that
\[\delta S=\delta S_{1}-\delta S_{0}=
\frac{\eta M}{\pi}(-\frac{1}{24}\rho^{-\frac{3}{2}}
+\rho^{-\frac{1}{2}})\times\]
\begin{equation}
\int_{0}^{\pi}d\theta\int_{0}^{2\pi}d\phi (\delta r).
\end{equation}
Then,
\begin{equation}
\delta S\vert_{S^{2}}=(-\frac{M}{24}+M)
\int_{0}^{2\pi}d\phi (\delta r),
\end{equation}
where $S^{2}$ denotes the two-sphere on $r=M$.

We evaluate variation $(\delta r)$ at $r=M$ by considering
null geodesics that cross each other at exactly
the surface $r=M$, going always
through original regions on the Kruskal diagrams for the two patches.
For such geodesics we have
$\bar{t}_{k=+1}=\bar{t}_{k=-1}$. Thus, if in passing
from the original regions of
patch $k=+1$ to the original regions of
patch $k=-1$ time $\bar{t}$
remains constant, time $t$ must change on $r=M$ according
to (Ref. Eqn. (4.4))
\begin{equation}
\left. \delta t \right|_{r=M}\equiv \left. t_{k=+1}-t_{k=-1} \right|_{r=M}=2M.
\end{equation}
In order to calculate the corresponding rate $\frac{dr}{dt}$ from patch
$k=+1$, we note that
\begin{equation}
\left. \frac{d\bar{t}}{dr} \right|_{r=M}
=\left. (\frac{dt}{dr}+\tan 2\alpha-\frac{1}{\cos 2\alpha})\right|_{r=M}=0.
\end{equation}
Hence,
\begin{equation}
\left. \frac{dt}{dr} \right|_{r=M}=-1.
\end{equation}
We have then
\begin{equation}
(\delta r)_{k=+1\rightarrow k=-1}=
\left. \delta t\right|_{r=M} \left. \frac{dr}{dt} \right|_{r=M}=-2M.
\end{equation}
Hence, from (A.6) we finally obtain
as the exact expression for the Euclidean action of a
black-hole pair for an observer in the asymptotic region
of patch $k=+1$
\begin{equation}
I=\delta S\vert_{r=M} = \frac{\pi}{6}M^{2}-4\pi M^{2}.
\end{equation}
The semiclassical production rate is then
\begin{equation}
e^{-I}\sim\exp(-\frac{\pi}{6}M^{2}+4\pi M^{2}).
\end{equation}
We interpret now the two factors in (A.12). 
For an asymptotic observer in patch $k=+1$,
the factor $e^{-\frac{\pi}{6}M^{2}}$ should give the full rate
of Schwarzschild black hole pair production in the gravitational
field created by a body with mass $M$ made of fermions in a
universe, and another body with the same
mass, but having exactly the antifermions to the fermions of
the first body, in the other universe. This factor should arise
from the nonfactorizability of the quantum state of a pair.
It would actually give the probability for such a state to
be factorizable, 
or in other words, the probability that
after cutting the full manifold at $r=A$ on any surface
$\bar{\Sigma}=\bar{\Sigma}^{+}\cup\bar{\Sigma}^{-}$, it is
divided into two topologically disconnected submanifols,
each for a single black hole in one coordinate patch.

The second factor in (A.12) gives the entropy
of the black hole, $S_{BH}=4\pi M^{2}
=\frac{1}{4}A_{BH}$, where $A_{BH}$ is the surface area of
a single black hole. This entropy can also be obtained from
the black hole temperature $T_{BH}$ by
insertion into the thermodynamic formula $T_{BH}^{-1}
=\frac{\partial S_{BH}}{\partial E}$, or by using the
instantonic procedure for single black holes described
in Sec. III.

Consistency of the above interpretation can only be achieved
if we look at the factor $e^{S_{BH}}$ as a
count of the number of physically
relevant black hole internal
states residing in between the event horizon and the interior
surface at $r=M$.
Positiveness of the full exponent
in (A.12) leads, on the other hand, to the remarkable feature
that although the rate of pair production is maximum for
Planck-sized black holes, once one of such pairs is formed,
the semiclassical probability (A.12) will tend to favour
processes in which the mass of the black holes
increases endlessly.

\pagebreak

\begin{center}
{\bf Legends for figures}
\end{center}

\noindent $\bullet$ Fig. 1: Kruskal diagrams for the two
coordinate patches ($k=\pm 1$) of the one-kink extended
Schwarzschild metric. Each of these patches is regarded as
providing the coordinates which describe a different universe.
Points on the diagrams represent 2-spheres. The null geodesic
discussed in the text is the straight line labelled
$a_{1}a_{2}a_{3}$ on the diagrams. The hyperbolae at $r=M$
are identified on, respectively, the original regions
($II_{+}$ and $I_{-}$) and the new regions ($III_{+}$ and
$IV_{-}$) created by the Kruskal extension.

\vspace{.5cm}

\noindent $\bullet$ Fig. 2: The half of the black-hole
instantons described in the text as depicted in the
coordinate patch $k=+1$. (a) Half of the Gibbons-Hawking
instanton. The Cauchy surface
$\Sigma^{+}=\Sigma_{p}^{+}\cup\Sigma_{u}^{+}$ is one boundary
and spatial infinity $\partial M_{\infty}^{+}$ is another.
$\Sigma^{+}$ represents an Einstein-Rosen bridge. The
amplitude $\langle\tau_{1}|\tau_{2}\rangle$ to go from
surface $\tau_{1}$ to the surface $\tau_{2}$ is given by
the action of the shaded sector. The Barvinsky-Frolov-Zelnikov
no-boundary wave function for the black-hole quantum state
discussed in Sec. IVB is defined on this half instanton.
(b) Half of the interior instanton, $0\leq r\leq 2M$,
discussed in Sec. III. The Cauchy surface $\tilde{\Sigma}^{+}
=\tilde{\Sigma}_{p}^{+}\cup\tilde{\Sigma}_{u}^{+}$ is one
boundary of this Euclidean manifold and the surface at the
horizon $\partial M_{h}^{+}$ is another. The amplitude
$\langle T_{2}|T_{1}\rangle$ to go from the surface $T_{1}$
to the surface $T_{2}$ is given by the action of the
corresponding shaded sector. If we were living in the
interior of a black hole with the mass of the universe,
the Hartle-Hawking no boundary wave function of the
universe could be defined on this half instanton. (c)
Half of the black-hole kink instanton discussed in Secs.
III, IV

\pagebreak

\noindent and V, covering both the exterior and the compact
interior region supporting the kink which is bounded at
$r=A$ from below. The disjoint union of the Cauchy
surface $\bar{\Sigma}^{+}=\tilde{\Sigma}^{+}\cup
\tilde{\Sigma}^{+}$ with the inner surface $\partial M_{A}^{+}$,
which is the maximal analytical extension of the
Einstein-Rosen bridge, is one boundary of this kinky
Euclidean manifold, and spatial infinity $\partial
M_{\infty}^{+}$ is another. The arguments for the wave
function $\Psi^{+}$ are the boundary values of the
quantum fields on the asymptotically flat part and on the
$r=A$ part of the analytically extended Einstein-Rosen
bridge $\bar{\Sigma}^{+}$. A similar half-instanton can
be constructed in patch $k=-1$. Identification of the
surfaces $\partial M_{A}$ of these two half-instantons
can only be made if a topology change is allowed to occur
by virtue of which a baby universe is branched off from
the kinky spacetime in such a way that the submanifolds
of the two patches are no longer topologically disconnected
to each other (see Sec. VB). In the figure, $\xi=2$,
$r_{h}=2M$ and $A=M$ for D=4, and $\xi=\frac{1}{2}$,
$r_{h}=M$ and $A=\frac{M}{\sqrt{2}}$ for D=5. The subscripts
$p$ and $u$ respectively denote physical and unphysical
regions.

\vspace{.5cm}

\noindent $\bullet$ Fig. 3: Penrose diagram of the
Schwarzschild black-hole kink. On the figure, surfaces at
$r=M$ with ends at $\iota^{+}$ in the two coordinate
patches are identified. Surfaces at $r=M$ with ends at
$\iota^{-}$ should also be identified on the two patches,
so that we have a unique and periodic conformal diagram. The
shaded interior regions correspond to imaginary values of
the Kruskal coordinates only for the tilded surfaces
$\tilde{\Sigma}$. A similar Penrose diagram can also be
constructed for the D=5 Tangherlini black-hole kink.
The meaning of the different symbols on the figure is
given in Secs. IV and V.

\end{document}